\documentclass[journal]{IEEEtran}

\ifCLASSINFOpdf

\else

\fi

\usepackage{amsmath}
\usepackage{threeparttable}

\usepackage{subfigure}							
\usepackage[pdftex]{graphicx}			
\usepackage{tikz}

\begin{document}

\title{Fast and Accurate Simulation Technique \\for Large Irregular Arrays}

\author{Ha Bui-Van,
				Jens Abraham,
				Michel Arts,
				Quentin Gueuning, Christopher Raucy
				\\David Gonzalez-Ovejero,~\IEEEmembership{Member,~IEEE},
        Eloy de Lera Acedo,
				and~Christophe~Craeye,~\IEEEmembership{Senior~Member,~IEEE}
\thanks{Ha~V.~Bui, Q. Gueuning, C. Raucy and C. Craeye are with the ICTEAM, Universit\'e catholique de Louvain, 1348 Louvain-la-Neuve, Belgium e-mail: (buivanha@uclouvain.be, christophe.craeye@uclouvain.be).}
\thanks{Jens Abraham and E.~de Lera Acedo are with Astrophysic Group, Cavendish Laboratory, University of Cambridge.}
\thanks{M. Arts is with ASTRON, the Netherlands.}
\thanks{ David Gonzalez-Ovejero is with the Institut d'\'Electronique et de T\'elecommunications de Rennes (IETR, UMR CNRS 6164), Universit\'e de Rennes 1,  35042 Rennes, France (david.gonzalez-ovejero@univ-rennes1.fr).}
\thanks{Manuscript received 2017}}

\markboth{IEEE Transactions on Antennas and Propagation, vol., no., 2017}%
{Ha V. Bui \MakeLowercase{\textit{et al.}}: Fast Simulation of Large Arrays}

\maketitle

\begin{abstract}

A fast full-wave simulation technique is presented for the analysis of large irregular planar arrays of identical 3-D metallic antennas. The solution method relies on the Macro Basis Functions (MBF) approach and an interpolatory technique to compute the interactions between MBFs. The Harmonic-polynomial (HARP) model is established for the near-field interactions in a modified system of coordinates. For extremely large arrays made of complex antennas, two approaches assuming a limited radius of influence for mutual coupling are considered: one is based on a sparse-matrix LU decomposition and the other one on a tessellation of the array in the form of overlapping sub-arrays. The computation of all embedded element patterns is sped up with the help of the non-uniform FFT algorithm. Extensive validations are shown for arrays of log-periodic antennas envisaged for the low-frequency SKA (Square Kilometer Array) radio-telescope. The analysis of SKA stations with such a large number of elements has not been treated yet in the literature. Validations include comparison with results obtained with commercial software and with experiments. The proposed method is particularly well suited to array synthesis, in which several orders of magnitude can be saved in terms of computation time.

\end{abstract}


\IEEEpeerreviewmaketitle


\section{Introduction}

\IEEEPARstart{M}{odern} radars~\cite{Dennis14}, massive MIMO~\cite{Larsson14} and radio astronomy arrays~\cite{Ellingson11,Eloy11} are in need of accurate modeling of their main electromagnetic properties. These are strongly impacted by mutual coupling and include the array impedance matrix and all embedded element patterns (EEP), i.e. the patterns of each individual elements in the presence of all other antennas. This can become a challenging task, knowing that arrays can consist of hundreds or thousands of  elements, as considered in the SKA project~\cite{SKA_pp,SKA} and in modern radar systems~\cite{Dennis14}. 

The arrays can be regular as well as non-regular. Irregular arrays exhibit a series of advantages over regular arrays, namely the randomization  of the side-lobes and of the mutual coupling effects~\cite{Nima12}. Both of these factors allow the operation of the arrays over a much larger frequency band compared to traditional regular arrays~\cite{Eloy15}, with a limited number of elements. 
The accurate knowledge of the antenna and array beams, as well as of the elements' impedance in the presence of mutual coupling~\cite{ChristopheRS11} is essential for the design of these systems, i.e. for the design of the antenna elements, the array configuration and the electronics connected to the antennas, especially the first stage amplification. The number of elements in the arrays, the geometrical complexity of the elements and the electrical size of these systems represent a challenge regarding their  computer modeling. In particular, the irregularity of the elements' positions in the array  prevents us from using certain approximations, such as periodic boundary conditions possibly combined with corrections for truncation~\cite{NetoTAP00}.
The present paper proposes a possible solution based on two novel contributions: (1) a fast full-wave technique able to accurately simulate large irregular arrays of complex structures in a matter of minutes; and (2) an analysis scheme that exploits this fast simulation technique for the analysis of extremely large arrays consisting of hundreds or thousands of elements.

Integral equation techniques solved with the method of moments (MoM) are very convenient for antenna problems, since they inherently account  for radiation. However, when one uses Gaussian elimination, the MoM solution has a complexity of $O( ({N_a} {N_e})^3)$, where ${N_a}$ is the number of antennas in the array, and ${N_e}$ is the number of elementary basis functions per element. 
For large arrays of complex antennas, the multilevel fast multipole (MLFMA) technique~\cite{Song95}, and its hybridization with other techniques (e.g. 
HOMoM-MLFMA~\cite{Zhao16}) are often exploited in an iterative scheme. 
Unfortunately, the calculation of the array impedance matrix and all EEPs in large finite arrays requires a new series of iterations for excitation at each antenna port.  
This often leads to an unaffordable computational effort for large arrays, besides the uncertainty related to the number of iterations.  A non-iterative technique is then favorable, provided that it can benefit from the proper compression of unknowns and from further accelerations. That is the case for techniques such as Macro Basis Functions (MBF)~\cite{Mosig2000}, Characteristic Basis Functions (CBF)~\cite{Mittra03,Maaskant08}, Synthetic Function Expansion (SFX)~\cite{Mate07,Mate09}, or entire/sub-entire-domain  (SED) basis functions~\cite{Lu04,Wang14}. 

MBF-type techniques achieve a compression of the MoM impedance matrix by replacing the original set of elementary basis functions by a new set of functions obtained through the solution of smaller problems. Once the MoM matrix size is reduced by means of the MBF technique, the $O((N_a N_e)^3)$ complexity for solving the system of equations is reduced to $O((N_a N_m)^3)$, where $N_m$ is the number of MBFs and $N_m<<N_e$. Nevertheless, the impedance matrix filling time remains $O((N_a N_e)^2)$, the same as for the original MoM, which is prohibitive for large arrays.
To overcome this obstacle, different techniques have been proposed, such as  the multipole expansion algorithm~\cite{Christophe06, DavidTAP13}, the interpolatory technique~\cite{DavidTAP11}, 
the perturbation technique based on the domain Green's function~\cite{Ludick14}, or the combination of multipole and Contour-FFT approaches for printed structures~\cite{Shambhu14}. The multipole expansion, however, suffers from the instability of near-field interaction, which appears when antennas are too close to each other, i.e. for distances smaller than about one wavelength. For dense wideband arrays, such as those envisaged for the SKA, antennas are very close to each other at low-frequencies, which requires additional efforts to explicitly calculate the near-field interactions. In this respect, the interpolatory technique presented in~\cite{DavidTAP11} is very effective in providing highly accurate solutions even in the near-field~\cite{QuentinEuCAP15}. Moreover, MBF interactions are rapidly evaluated from a model, which is built beforehand and is independent of the array configuration. The method, renamed later as HARP~\cite{QuentinAPS15}, has shown great capability and has been extended to the analysis of antennas made of wires, in order to model the SKA1 - Low arrays~\cite{Eloy15,Eloy12}. 

Preliminary results of HARP on SKA1-Low antennas have been presented in recent conference papers~\cite{QuentinEuCAP15, QuentinAPS15,HaEuCAP16}. 
Nonetheless, limited details were presented about the technique itself and its performance.
The present paper further extends this method including a detailed description of the proposed technique, new improvements and extended validation. 
HARP is now capable of simulating SKA stations over the whole frequency band (50\,MHz to 350\,MHz) in a matter of minutes for each frequency point.

Additionally, two approaches are presented to exploit the fast simulation method for the analysis of extremely large arrays, namely, the sparse matrix approach and the tessellation approach. These methods make use of the so-called radius of influence (RI) concept, which is often applied to the analysis of large arrays~\cite{EloyICEAA11}. 
The first approach calculates only the interaction between antennas whose  relative distance is smaller than the RI, which leads to a sparse interaction matrix that can be effectively LU-decomposed and solved exploiting sparse-matrix libraries.
In the tessellation approach, the full array is first partitioned into overlapping subarrays. 
The EEPs are retained only for the internal (or non-overlapping) part of the subarray.
The size of the internal part and of the extended surrounding area of the subarray are determined so as to minimize the computational cost. As each subarray can be analyzed independently, this approach enables the simulation of very large arrays of complex elements on a normal desktop computer; it also facilitates parallel implementation. 
Finally, as explained above, the MBF approach allows the fast calculation of induced currents for excitation at each port. The calculation of corresponding EEPs is then expressed as  superposition of pattern multiplication problems, one per MBF~\cite{Christophe09}, and the array factors calculation is accelerated using the non-uniform FFT (NFFT)~\cite{Christophe09} -- \cite{Capozzoli10}.

\begin{figure}[!thb]\centering
\includegraphics[scale=0.8,clip,trim={2cm 0cm 2cm 2.5cm}]{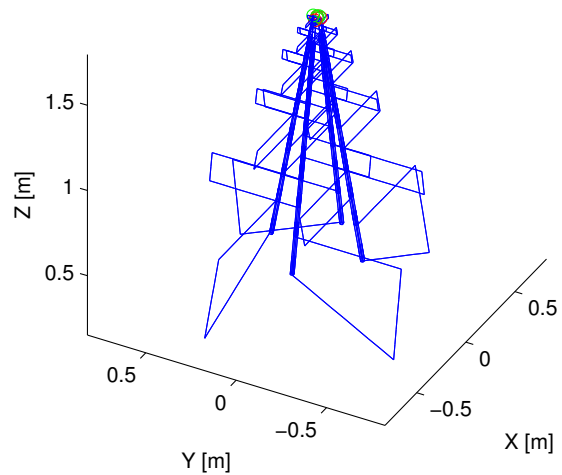}
\caption{Wire mesh of the SKA Log-periodic Antenna (SKALA) having a footprint of 1.2$\times$1.2 $\rm{m^2}$. The SKALA consists of 4 arms supported by 4 spines, with two orthogonal feeds (on top), providing two polarizations. }
\label{fig:meshSKA}
\end{figure}

\begin{figure}[!htb]\centering
\includegraphics[scale=0.65,clip,trim={0cm 0cm 0cm 0.5cm}]{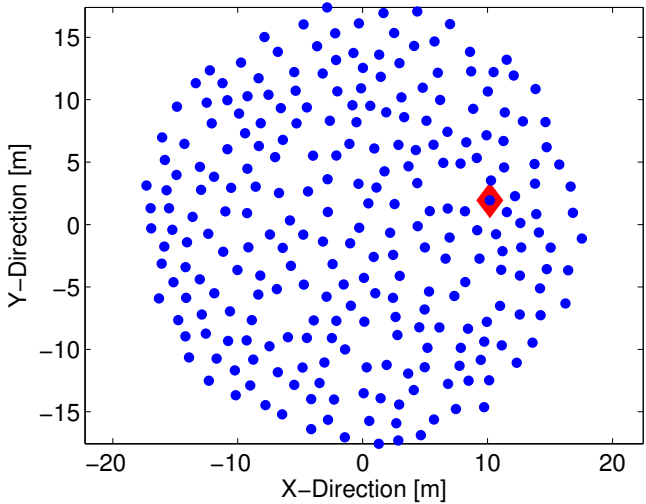}
\caption{Layout of a SKA1-low station consisting of 256 SKALA antennas with a minimum distance of 1.35\,m.}
\label{fig:array256}
\end{figure}

In this paper, the SKA1-Low arrays are taken as a case study to extensively validate the performance of HARP. Upon its completion, the SKA1-Low instrument  will consist of 512 stations, each with 256 log-periodic antennas (SKALA)~\cite{Eloy15,Eloy12}, (see Fig.~\ref{fig:meshSKA}), arranged in a pseudo-random configuration, as shown in Fig.~\ref{fig:array256}.
 With an ultra-wideband characteristic (7:1 fractional bandwidth) and its complex structure, the  modeling and simulation of the SKA1-Low frequency array presents a challenge as the array is quite dense at low frequencies, while at high-frequencies, the antenna needs to be meshed precisely. The accurate modeling of the SKALA antenna has a strong impact on the results obtained with a given simulation software. The results obtained using HARP will be validated with those calculated using commercial software, and with measurements of the isolated elements and a small 3$\times$3 array.

The remainder of the paper is organized as follows. In Section~\ref{sec:technique}, the full-wave technique and improvements made to HARP are explained. The sparse matrix and tessellation approaches for extremely large arrays are detailed as well.  In Section~\ref{sec:numerical}, numerical results on the SKA1-Low case are presented. Data obtained using HARP are compared with those from CST Studio~\cite{cst}, WIPL-D~\cite{wipld}, and measurements for some occasions. In Section~\ref{sec:largeArray}, the performance of the tessellation scheme on very large arrays   is examined and compared with those from the traditional RI approach on different arrays of SKALA antennas. Finally, conclusions  are drawn in Section~\ref{sec:conclusion}.

\section{Fast Full-wave Technique}  \label{sec:technique}

The full-wave approach proposed in this paper is based on the interpolatory method presented in~\cite{DavidTAP11}, renamed as HARP in~\cite{QuentinAPS15}, which uses an interpolatory approach to effectively deal with the analysis of metallic antenna arrays in free space. In this section, we will explain in detail the HARP method, and present improvements that make HARP able to provide with the properties of complex radiators, such as the SKALA antenna.  The tessellation approach, which exploits HARP, is also introduced; it enables the effective simulation of extremely large arrays. For the sake of clarity, the interpolatory method is first recalled.  Interested readers are referred to~\cite{DavidTAP11} for detailed explanations regarding the interpolatory technique. In the present work, the MBFs are obtained using the ``primary-and-secondaries'' approach~\cite{Mittra03}, where the primary MBF corresponds to the current distribution on an isolated element and the secondary MBFs are the currents induced on a set of neighboring antennas by the element that supports the primary MBF.

\subsection{Background of HARP}

The interpolatory approach was proposed in~\cite{DavidTAP11} to quickly compute the interaction matrix between MBFs, which are the entries of the reduced MoM matrix. 
The interaction between a source MBF $S$ on antenna $n$ placed at the origin and a testing MBF $T$ on antenna $m$, positioned at $(r_{mn},\hat{\alpha})$, was first approximated as
\begin{equation}
Z_{\rm{TS}}^{\rm{app}}(r_{mn},\hat{\alpha}) \approx -j\omega \mu ~\vec{F}_{T,m}^{\hat{\alpha}, *}~.~\vec{F}_{S,n}^{\hat{\alpha}}~\frac{e^{-jkr_{mn}}}{4\pi r_{mn}}
\end{equation}
where $\hat{\alpha}$ points from the reference point in the source antenna to the reference point in the testing antenna and $r_{mn}$ represents the inter-element distance, $\omega$ is the angular frequency, $\mu$ is  the free-space permeability, $\vec{F}_{T,m}^{\hat{\alpha}}, \vec{F}_{S,n}^{\hat{\alpha}}$ are the radiation patterns of the testing and source MBFs, respectively, and ($*$) stands for complex conjugate.  The actual interaction is then obtained as a sum of the above far-field approximation and a complementary part, which is represented using a harmonic-polynomial model. The model is built by firstly calculating explicitly the exact interaction between MBFs over a grid with a limited number of points. A mathematical representation of the difference between the exact and far-field approximated interaction is then constructed after phase correction and an appropriate change of variables. The first step can be expressed in a compact form as 
\begin{equation}
B_{\rm{TS}}(r_{mn},\hat{\alpha}) = \frac{Z_{\rm{TS}}^{\rm{exact}}(r_{mn},\hat{\alpha}) - Z_{\rm{TS}}^{\rm{app}}(r_{mn},\hat{\alpha})}{e^{-jkr_{mn}}} \label{eq:step12} 
\end{equation}
where $Z_{\rm{TS}}^{\rm{exact}}(r_{mn},\hat{\alpha})$ is the exact interaction explicitly calculated on a pre-defined polar-radial grid. 
The result of Eq.~(~\ref{eq:step12}) is then modeled in a  harmonic-polynomial (HARP) form~\cite{DavidTAP11} with a proper change of variables, i.e. $d = 1/r^2$:
\begin{equation}
B_{\rm{TS}}(d,\hat{\alpha}) = \sum_{p=-P}^P e^{jp\alpha} \sum_{q=0}^Q c_{pq}~d^{q} \label{eq:repHARP}
\end{equation}
where $P$ and $Q$ are the orders of Fourier Series and polynomial, respectively, and $c_{pq}$ are coefficients calculated in the least-squares sense. Once the model has been built, the interaction between MBFs  is rapidly calculated by adding up the far-field contribution and the complementary part evaluated from the model. 
\begin{equation}
Z_{\rm{TS}}(r_{mn},\hat{\alpha}) = Z_{\rm{TS}}^{\rm{app}}(r_{mn},\hat{\alpha}) + B_{\rm{TS}}(r_{mn},\hat{\alpha})e^{-jkr_{mn}} \label{eq:HARP}
\end{equation}

\subsection{Improved HARP model}
Improvements have been introduced to the original version of HARP to capture the interaction of complex structures such as the SKALA element. The first enhancement appears at the change of variables step, where $d = 1/r$ is now implemented instead of $d = 1/r^2$ as in Eq.~(\ref{eq:repHARP}). This modification is physically interpreted as taking into account not only the contribution of $1/r^2$, but also of $1/r^3$ field components~\cite{Amir02}, which are non-negligible in the case of SKALA. Such a change of variables is also consistent with the multipole-based model in~\cite{QuentinEuCAP15}, where a Laurent series versus the $r$ variable  is obtained. The latter work, however, suffers from the stability issues in the very near field. Therefore, in this case, we followed~\cite{DavidTAP11} to build the HARP model using the new change of variables.

Additionally, improvements have been added to the implementation of HARP. The DFT approach has been used instead of a least-squares estimator to obtain the harmonic coefficients ($c_{pq}$). Also, the sampling points on the grid are determined by an appropriate sampling routine, i.e. a regular grid on angular variable, and inversely regular along the radial distance (regular in $1/r$ scale). As it will be shown in Section~\ref{sec:numerical}, HARP is now able to accurately model the interaction between the MBFs of the SKALA over the whole frequency band, while satisfying the minimum distance between elements in the SKA-station.

Regarding the computational aspect using HARP, the complexities for filling the reduced matrix and solving the system of equations are $O(N_a^2N_{m}^2S)$ and $O((N_{a}N_{m})^3)$, respectively, where $S$ is a small factor related to the calculation of Eq.~(\ref{eq:HARP})~\cite{DavidTAP11}. 
The number of MBFs is generally  smaller by about two orders of magnitude than the number of basis functions used to describe the current on the antenna (related to meshing); the drastic reduction in the number of unknowns enables the acceleration of the analysis using HARP. 
In~\cite{DavidTAP11}, it is shown that interpolatory (or HARP) technique is more efficient than the multipole expansion~\cite{Christophe06}. Moreover, HARP is able to accurately evaluate the near-field interaction for closely packed arrays.

\subsection{Pattern Calculations}
The use of MBFs allows one to write the pattern of each antenna as a sum of MBF patterns multiplied by the corresponding coefficients for the MBF. Hence, the array pattern can be expressed as:
\begin{align}
\vec{F}_{array}(\theta,\phi) &= \sum_{i=1}^{N_{a}}\sum_{j = 1}^{N_{m}} c_{ij}\,\vec{F}_j(\theta,\phi) \,e^{jk(u_xx_i + u_yy_i)}\\
 &= \sum_{j = 1}^{N_{m}} \vec{F}_j(\theta,\phi) \sum_{i=1}^{N_{a}} c_{ij}\,e^{jk(u_xx_i + u_yy_i)} \label{eq:arrayPat}
\end{align}
where $(x_i,y_i)$ is the position of the $i^{\rm{th}}$ antenna, $\hat{u} = (u_x,u_y,u_z)$ is a unit vector in the direction $(\theta,\phi)$ of interest, $\vec{F}_j(\theta,\phi)$ is the pattern of MBF $j$, and $c_{ij}$ is the coefficient for MBF $j$ on antenna $i$. The swapping of the summations is possible since only one set of MBFs is used for all-identical-antennas of the array. As the MBF patterns in Eq.~(\ref{eq:arrayPat}) are factored out, the second summation becomes the array factor, which can be rapidly evaluated using the N-FFT~\cite{Dutt93,Christophe09,Capozzoli10} for irregular arrays. 
It is noted that since the MBF patterns are smooth functions, they are evaluated over a much sparser grid in advance. The higher resolution patterns then can be rapidly obtained using 2-D interpolation on the grid used for the N-FFT.
This approach is also applied for the calculation of each EEP of the array, where the induced currents are quickly calculated for excitation at a given antenna port.

\begin{figure}[!htb]\centering
\includegraphics[scale=0.65,clip,trim={0cm 0cm 0cm 0cm}]{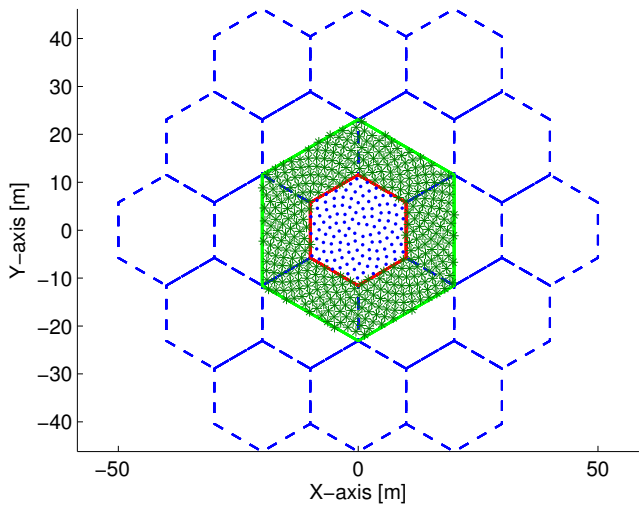}
\caption{Tessellation (subarray) approach: inner non-overlapping part (blue dashed hexagonal) and an extended area limited to the green hexagonal.}
\label{fig:subarray}
\end{figure}

\subsection{Analysis of Very Large Arrays}  \label{subsec:LargeArray}
For extremely large arrays, comprising thousands of antennas, the direct implementation of HARP on the whole array would require too much memory and computation time. An effective approach exploiting HARP for such arrays is then needed.
Generally, the so-called radius of influence (RI) concept is often applied~\cite{EloyICEAA11, Ludick14} for the analysis of very large arrays; here the interactions are only considered when the separation between elements is smaller than the RI. This results in a sparse  matrix for the reduced MoM system of equations. 
The filling time of this sparse matrix and the solution time of the system of equations are significantly reduced owing to the use of sparse operations. 
More precisely, sparse matrix libraries~\cite{sparseMatlab} are applied to obtain a sparse LU-decomposition of the matrix. Fast solutions can then be computed for any excitation law.
The approach, however, requires a large RI~\cite{EloyICEAA11} in order to obtain a highly accurate pattern, which implies an increase of the computational cost.

An alternative technique is proposed below; it relies on the tessellation of the array into subarrays. The RI concept is then applied to each subarray individually. The idea is illustrated in Fig.~\ref{fig:subarray}, where the elements in the array are clustered into subarrays, i.e. hexagonal shapes with the inner non-overlapping part of radius $\rm{R_{in}}$ (blue dashed line), and an extended area between the inner radius $\rm{R_{in}}$ and the outer one $\rm{R_{in}} + \rm{R_{ext}}$ (green area).
The EEPs in the array are obtained by analyzing separately each subarray including the inner and the outer parts, and retaining only the EEPs of elements within the inner part. The tessellation approach deals with subarrays having much smaller size  than the whole array, which will ease the computational effort of solving the MoM system of equations. 
The use of an outer part with a width of $R_{\rm{ext}}$ at least equal to $R_{\rm{in}}$ helps to increase the accuracy of the results.

An optimal choice of $\rm{R_{ext}}$ can be made by analyzing the computational complexity of the approach. Assuming that a random array has a constant antenna density, the number of antennas on a subarray (inner and outer parts) will be proportional to $\rm{(R_{in}+R_{ext})^2}$. The number of subarrays will be of order of $N_a/\rm{{R_{in}^2}}$. Following the calculation in~\cite{DavidTAP11}, the computational effort for filling the matrix can be expressed as a function of $\rm{R_{in}}$:
\begin{equation}
C = \alpha\frac{N_a}{\rm{R_{in}}^2}(\rm{R_{in}} + \rm{R_{ext}})^4  \label{eq:complexity}
\end{equation}
where $\alpha$ is a constant taking into account all other constant factors such as the number of MBFs. $C$ presents a minimum when $\rm{R_{in} = R_{ext}}$. The same minimum can be found regarding the complexity of solving the system of equations. This gives a rule of thumb for the choice of  $\rm{R_{in}}$ and $\rm{R_{ext}}$ so as to optimize the quality of the solution while minimizing the computational complexity.

One of the important advantages of HARP is the re-usability of the model, which is established only once and can be exploited to analyze any array made of the same element. This feature is very helpful for the tessellation approach, as in this scenario, several subarrays can be simulated separately, which straightforwardly allows parallelization. The implementation and evaluation of HARP using the sparse matrix and tessellation approaches will be studied in Section~\ref{sec:largeArray}.

\begin{figure}[!htb]\centering
\includegraphics[scale=0.65,clip,trim={2cm 0cm 2cm 2cm},height = 6.5cm,width = 9cm]{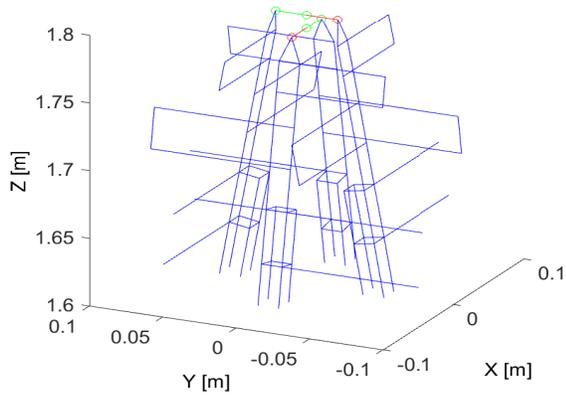}
\caption{Zoom in of the  upper part of the SKALA mesh implemented in this work: 2 feed points on top and 4 cages at the bottom.}
\label{fig:meshSKA_upper}
\end{figure}

\section{Numerical and Experimental Validation}   \label{sec:numerical}
Preliminary data at a single frequency has been presented in recent conference papers~\cite{QuentinAPS15}, \cite{HaEuCAP16} with limited results on relatively small arrays and on a SKA1-Low station. In the present paper, the performance of HARP is studied in depth on several examples, which include an isolated antenna, a small compact array, and the SKA station, shown in Fig.~\ref{fig:array256}, at  50\,MHz, 110\,MHz, 200\,MHz and 350\,MHz, covering the whole SKA1-Low band. Simulated results from two commercial software, CST Microwave Studio~\cite{cst} and WIPL-D~\cite{wipld}, are taken as a reference to examine the performance of HARP, while measured data is presented whenever they are available. Different parameters are analyzed, including input impedance, coupling coefficients, EEPs and array patterns. In all cases, the data for one polarization is shown, while the results for the  other polarization have the same quality and are not shown. The SKALA model used in HARP consists in a wire mesh comprising 1218 basis functions. 
This HARP implementation is based on a MoM code that considers wires, with the Poklington approximation~\cite{Balanis89}. The 4 cm diameter tube that supports each arm does not satisfy that approximation in the upper part of the frequency range. Therefore, those tubes have been replaced by wire cages with a square cross-section and a perimeter equal to that of the tube. Fig.~\ref{fig:meshSKA} shows a global view of the structure, while Fig.~\ref{fig:meshSKA_upper} shows a zoom near the feed points and near the cage.
 The numbers of MBFs are 15, 20, 45, 70 at 50\,MHz, 110\,MHz, 200\,MHz and 350\,MHz, respectively (linearly scaled versus frequency).

\begin{figure}
\centering
\includegraphics[scale=0.6,clip,trim={0cm 0cm 0cm 0cm},height = 6.5cm,width = 9cm]{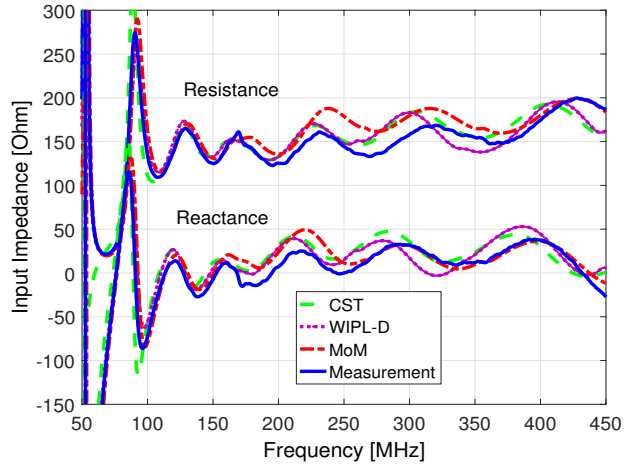}
\caption{Real and imaginary parts of the self-impedance of SKALA. Measurement with antenna standing on a 2\,m square metallic ground.}
\label{fig:Z11}
\end{figure}

\begin{figure}
\centering
\includegraphics[scale=0.5,clip,trim={0cm 0cm 0cm 2cm},height =6cm,width = 9cm]{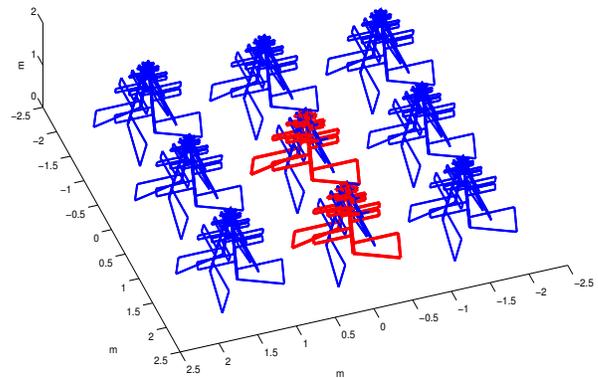}
\caption{A compact array of 3$\times$3 SKALA antennas with spacing of 1.5\,m.}
\label{fig:array9}
\end{figure}

\begin{figure}
\centering
\includegraphics[scale=0.5,clip,trim={0cm 0cm 0cm 0cm},height = 6.5cm,width = 9cm]{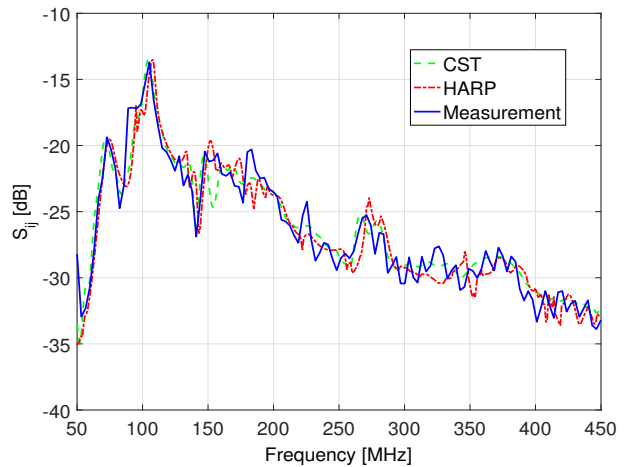}
\caption{Coupling coefficient in a 3$\times$3 array, between center element and middle of edge shown in Fig.~\ref{fig:array9}.}
\label{fig:Sij}
\end{figure}

\begin{figure}[!htb]\centering

\subfigure[Real Part at 50\,MHz.]{\includegraphics[scale=0.235,clip,trim={0.5cm 0cm 1.25cm 0.5cm}]{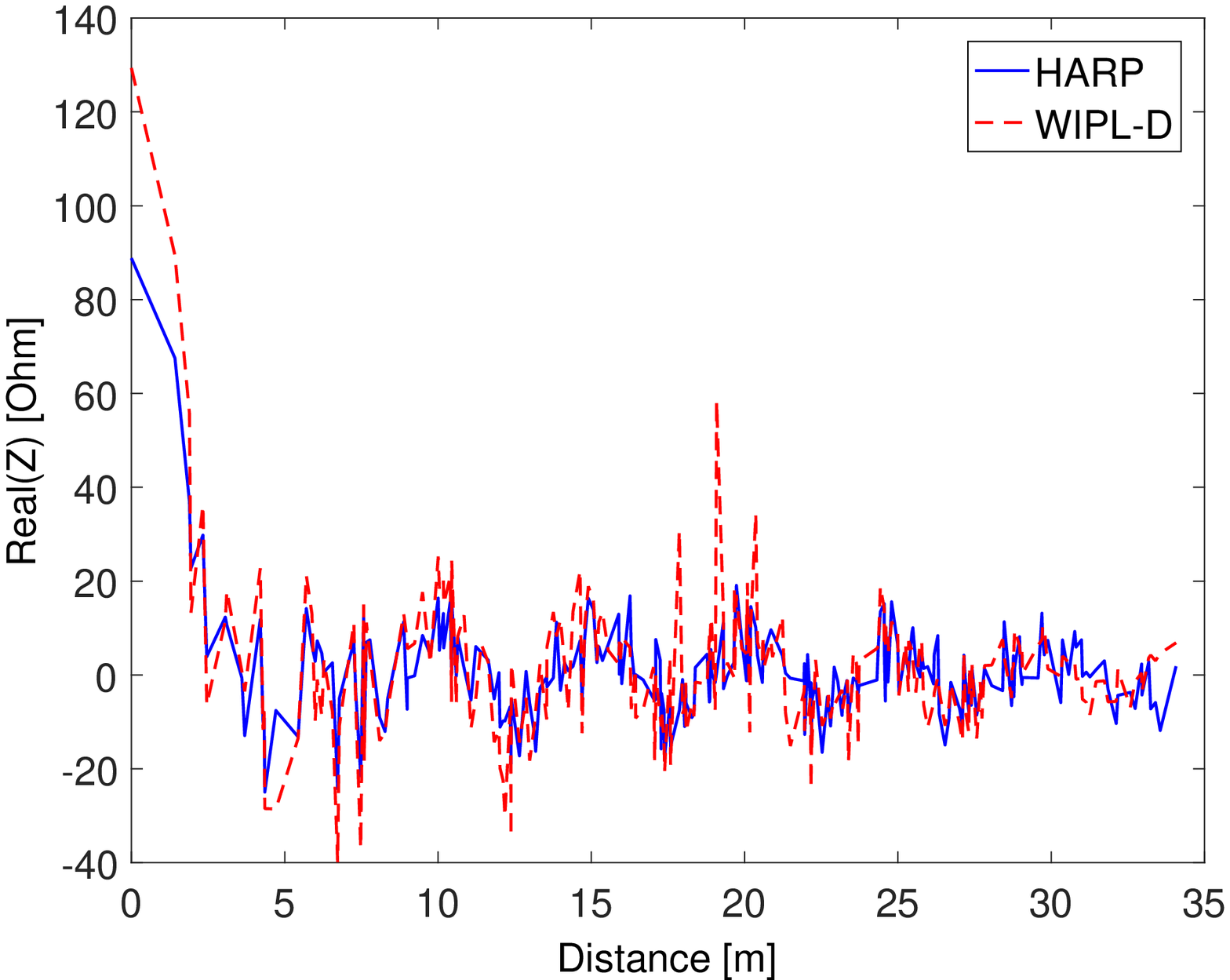}}
\subfigure[Imaginary Part at 50\,MHz]{\includegraphics[scale=0.235,clip,trim={0.5cm 0cm 1.25cm 0.5cm}]{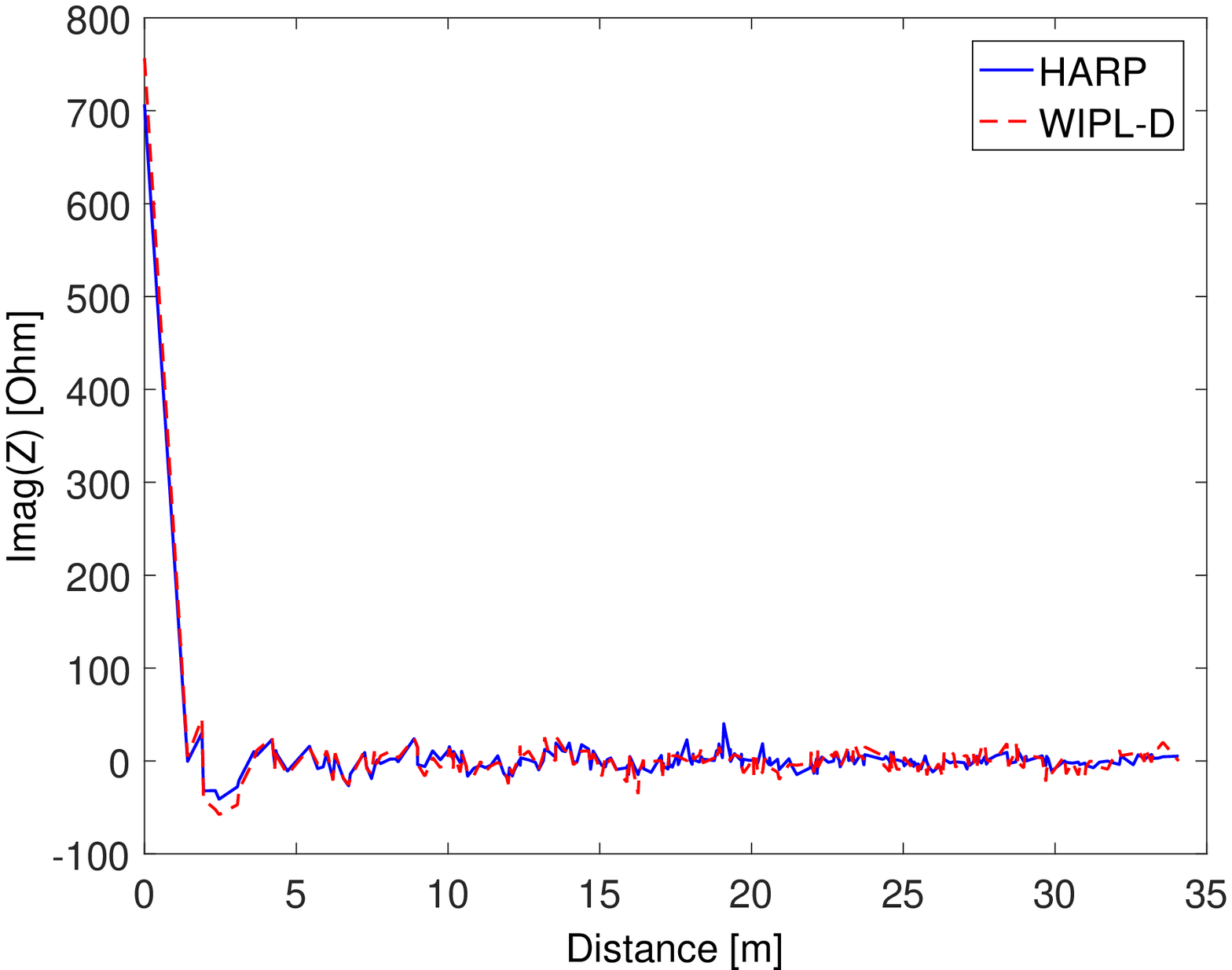}}

\subfigure[Real Part at 110\,MHz.]{\includegraphics[scale=0.235,clip,trim={0.5cm 0cm 1.25cm 0.5cm}]{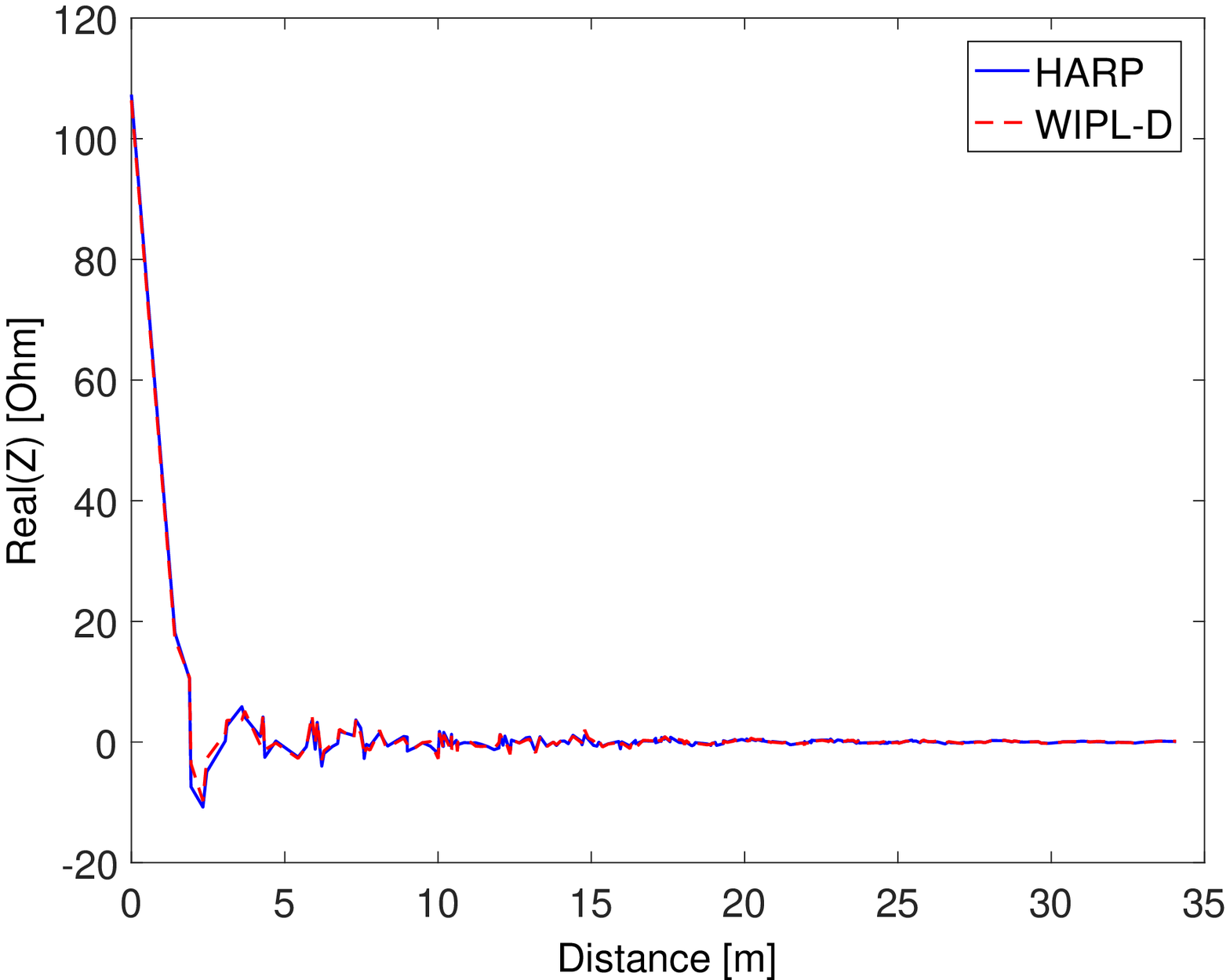}}
\subfigure[Imaginary Part at 110\,MHz]{\includegraphics[scale=0.235,clip,trim={0.5cm 0cm 1.25cm 0.5cm}]{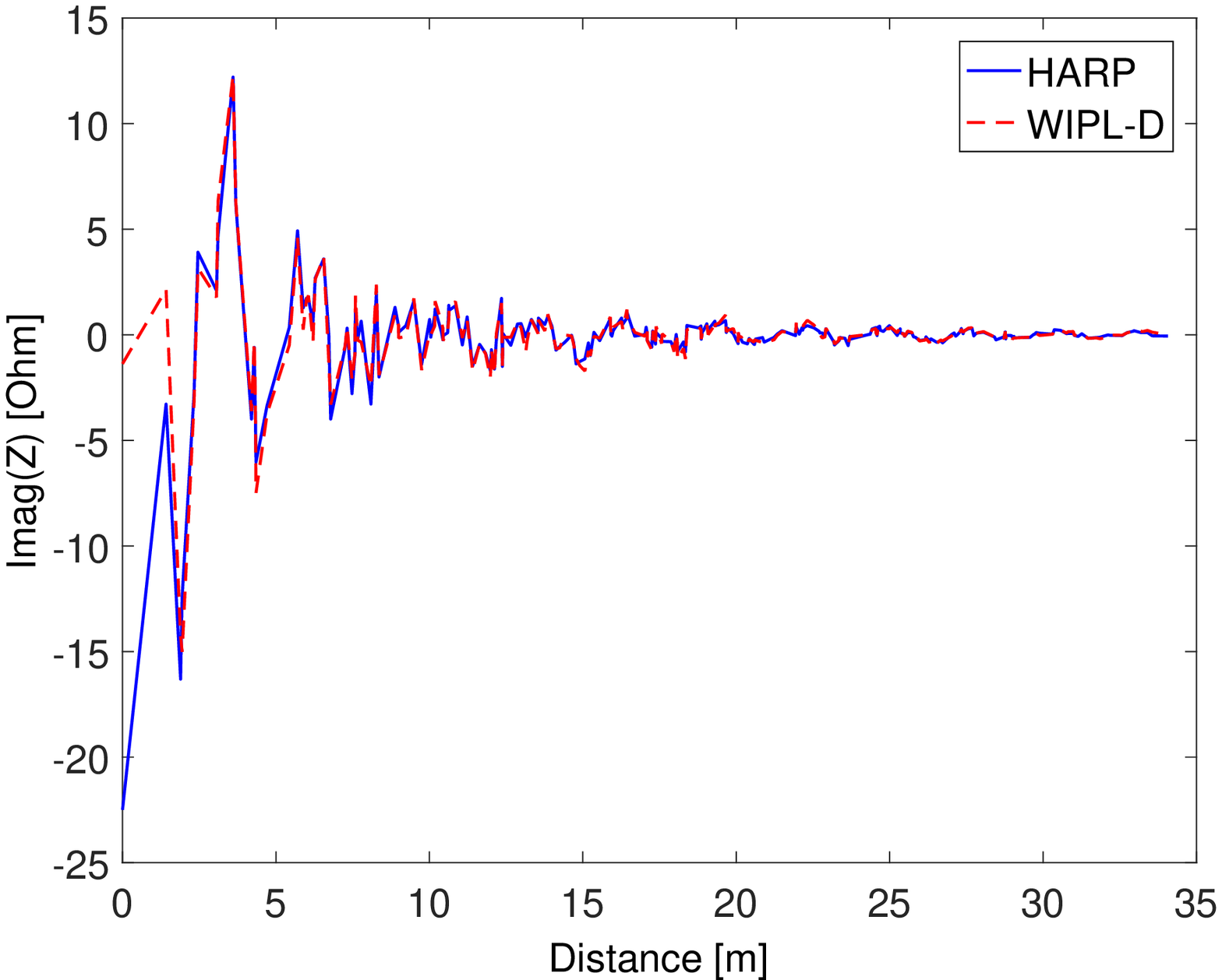}}

\subfigure[Real Part at 200\,MHz.]{\includegraphics[scale=0.235,clip,trim={0.5cm 0cm 1.25cm 0.5cm}]{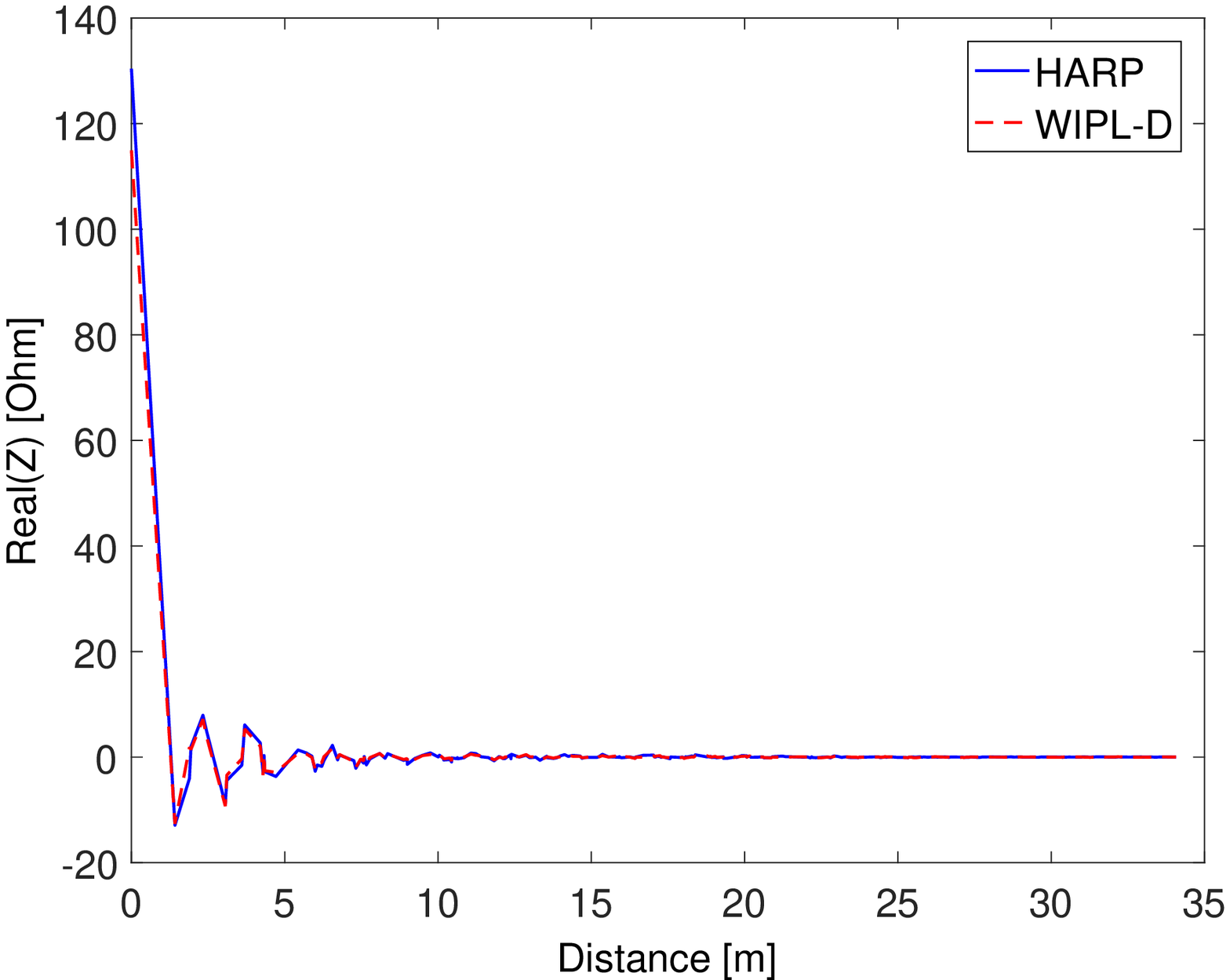}}
\subfigure[Imaginary Part at 200\,MHz]{\includegraphics[scale=0.235,clip,trim={0.5cm 0cm 1.25cm 0.5cm}]{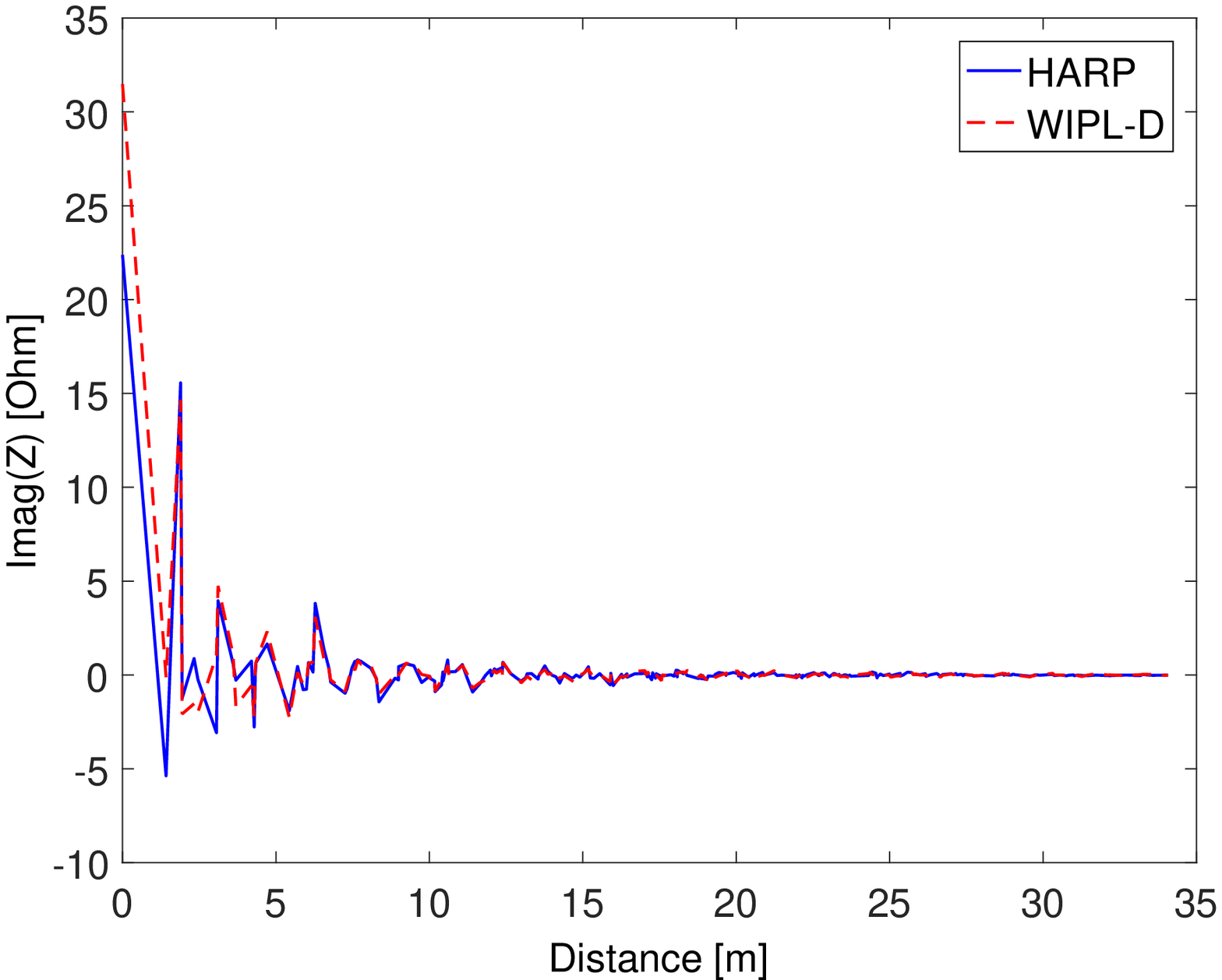}}

\subfigure[Real Part at 350\,MHz.]{\includegraphics[scale=0.235,clip,trim={0.5cm 0cm 1.25cm 0.5cm}]{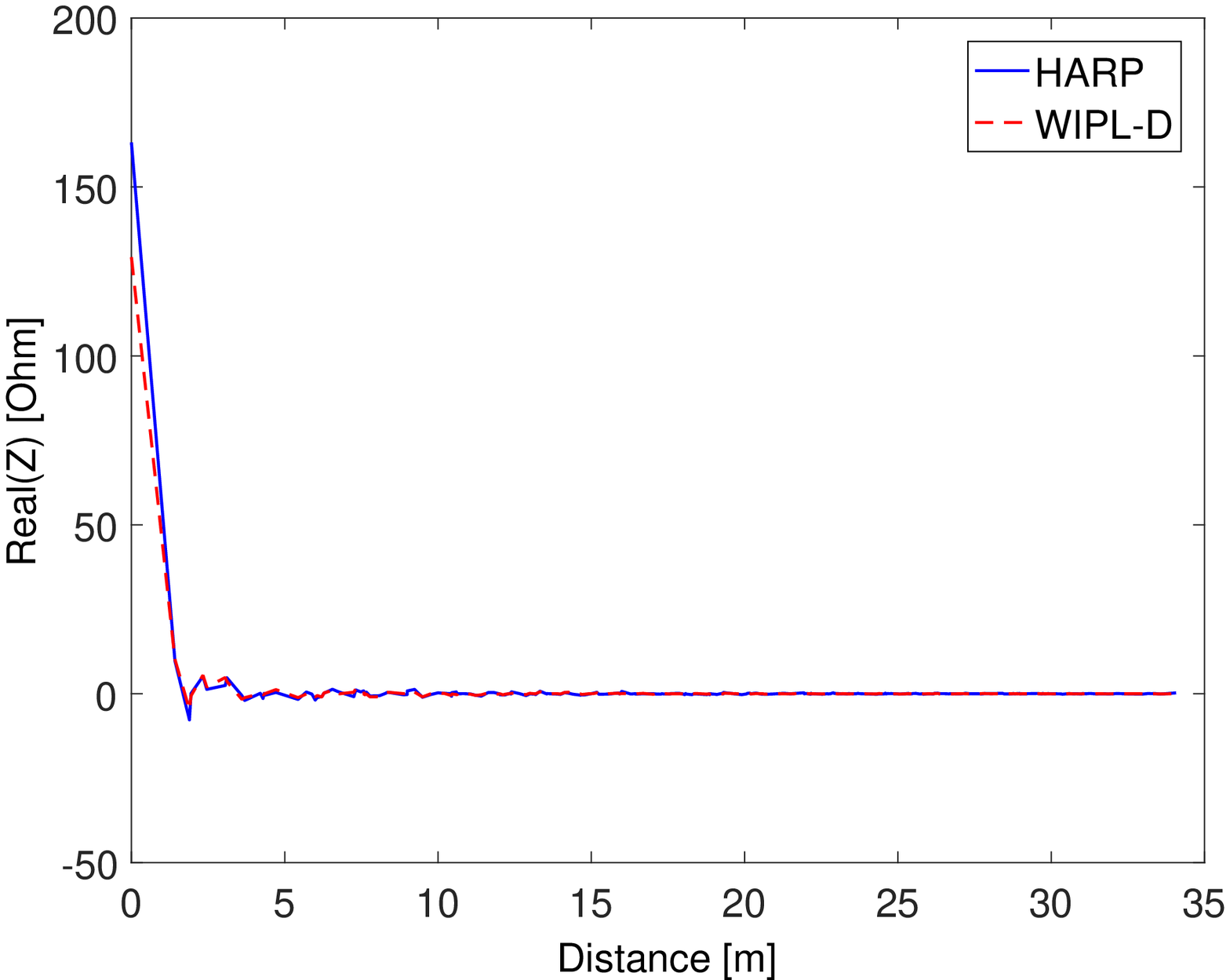}}
\subfigure[Imaginary Part at 350\,MHz]{\includegraphics[scale=0.235,clip,trim={0.5cm 0cm 1.25cm 0.5cm}]{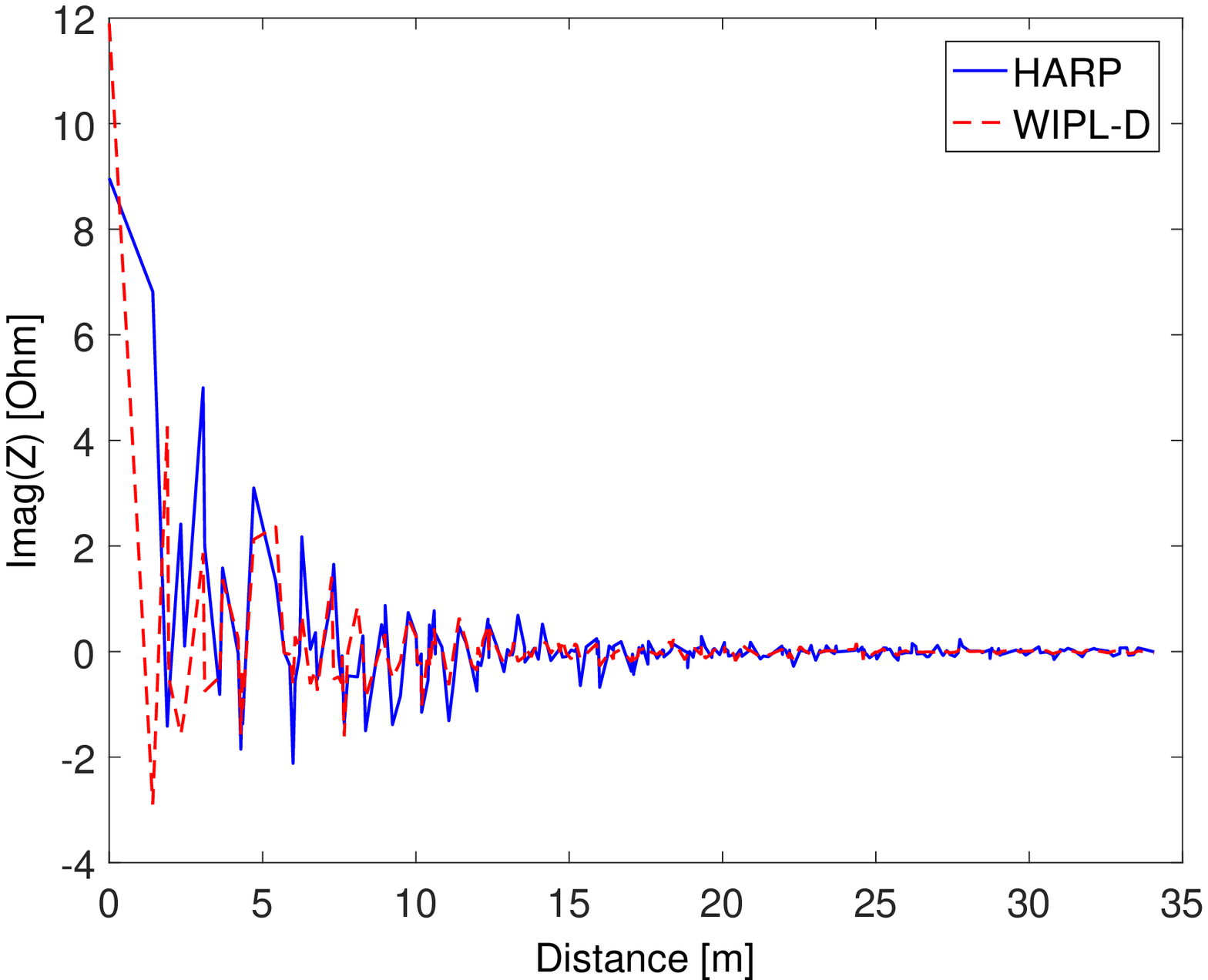}}

\caption{Mutual impedance of a random antenna ${i}$ in the SKA station (a red dot in Fig.~\ref{fig:array256}) and other antennas ${j}$ sorted by the distance to antenna ${i}$.}
\label{fig:MutualImpedance}

\end{figure}

\subsection{Self and Mutual Impedance}
Due to the complexity of the design, an accurate model of the SKALA in Fig.~\ref{fig:meshSKA} is critical for analyzing the performance of the array, e.g. self and mutual impedances are required to study parameters such as noise coupling~\cite{CCTAP04,IvashinaTAP10}.  The input impedance is then the first parameter to study.  Fig.~\ref{fig:Z11} shows the self-impedance simulated using different software, along with the measured data over the low-frequency band. A small deviation of the order of $20\,\Omega$ is seen around 200--300\,MHz between the simulations and the measurements. This can be attributed to the fact that different software packages model and mesh the SKALA differently, and that the SKALA in the measurement had a metallic ground limited to 2\,m, whereas the simulations assume an infinite ground plane. These variations may also explain some of the differences among patterns in  the following sections.

Another validation example carries on a 3$\times$3 regular array of SKALAs with element spacing of 1.5\,m; as shown in Fig.~\ref{fig:array9}. The coupling coefficient is calculated between the center antenna and another element at the middle of the edge, as shown in Fig.~\ref{fig:Sij}. This example is a repetition of an example in~\cite{QuentinAPS15} with an improved version of HARP used here. A clear improvement is seen for HARP w.r.t. the result in~\cite{QuentinAPS15} as the HARP and CST simulations follow well the measurement across the whole band. 

The mutual impedance is studied for the SKA station as well. Fig.~\ref{fig:MutualImpedance} plots the real and imaginary parts of the mutual impedance $Z_{\rm{ij}}$ for different frequencies  in a SKA station between an arbitrary antenna ${i}$ (a red dot in Fig.~\ref{fig:array256}) and another antenna ${j}$ running from 1 to 256; the obtained values are sorted versus distance to antenna ${i}$. A good agreement is found between HARP and WIPL-D for coupling between antennas. The main difference regards the self-impedance, i.e. zero distance, which is also illustrated in Fig.~\ref{fig:Z11}. The results confirm the capability of the current version of HARP to capture the interaction between SKALA antennas in their array environment.

\begin{figure}[!htb]\centering
\centering
\includegraphics[scale=0.45]{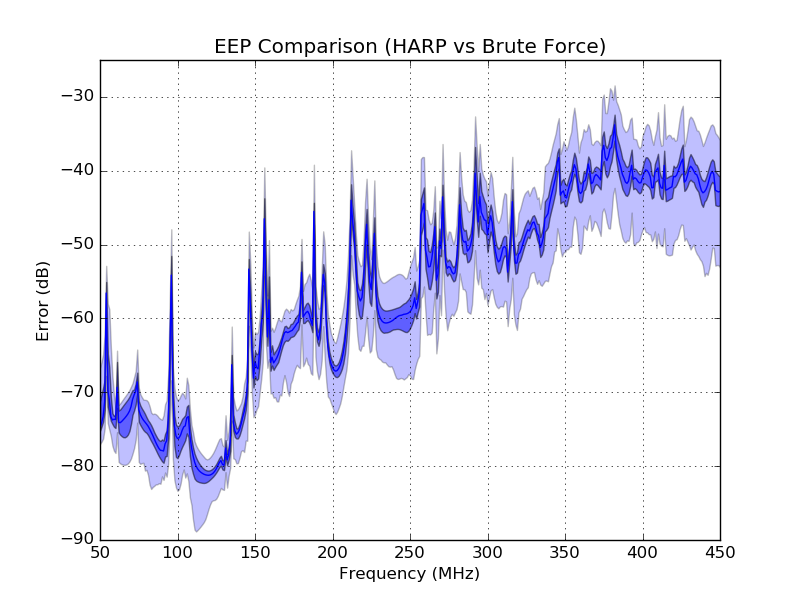}
\caption{Error of the EEPs between the HARP and the brute force solution over all 16 antennas in all directions against frequency. The median, 25 \% and 75 \% percentiles as well as the minimum and maximum error are shown by the blue line, the dark blue shaded and the light blue shaded areas, respectively.}\label{fig:eep_error}
\end{figure}

\begin{figure}[!htb]\centering

\subfigure[50\,MHz.]{\includegraphics[scale=0.235,clip,trim={1.25cm 0cm 1cm 0cm}]{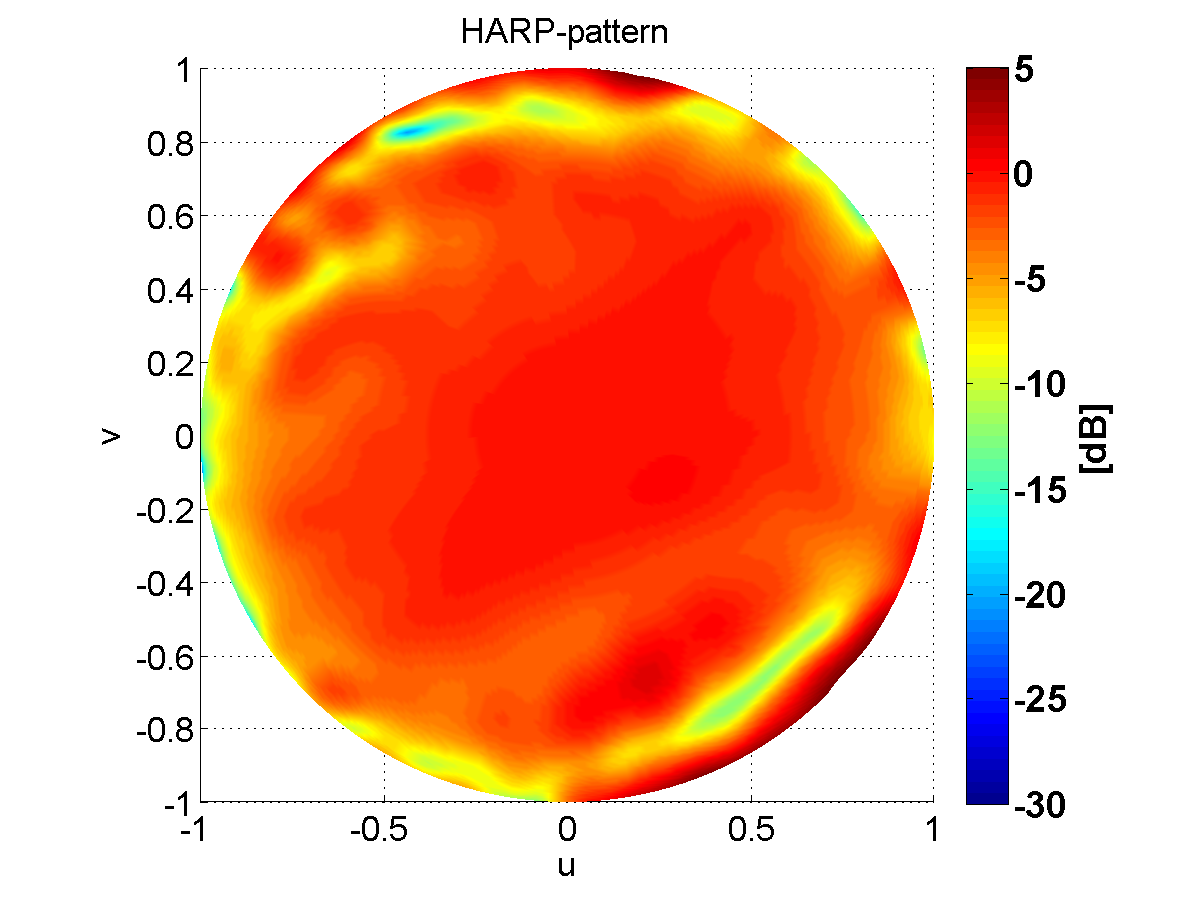}}
\subfigure[Error Pattern 50\,MHz]{\includegraphics[scale=0.235,clip,trim={1.25cm 0cm 1cm 0cm}]{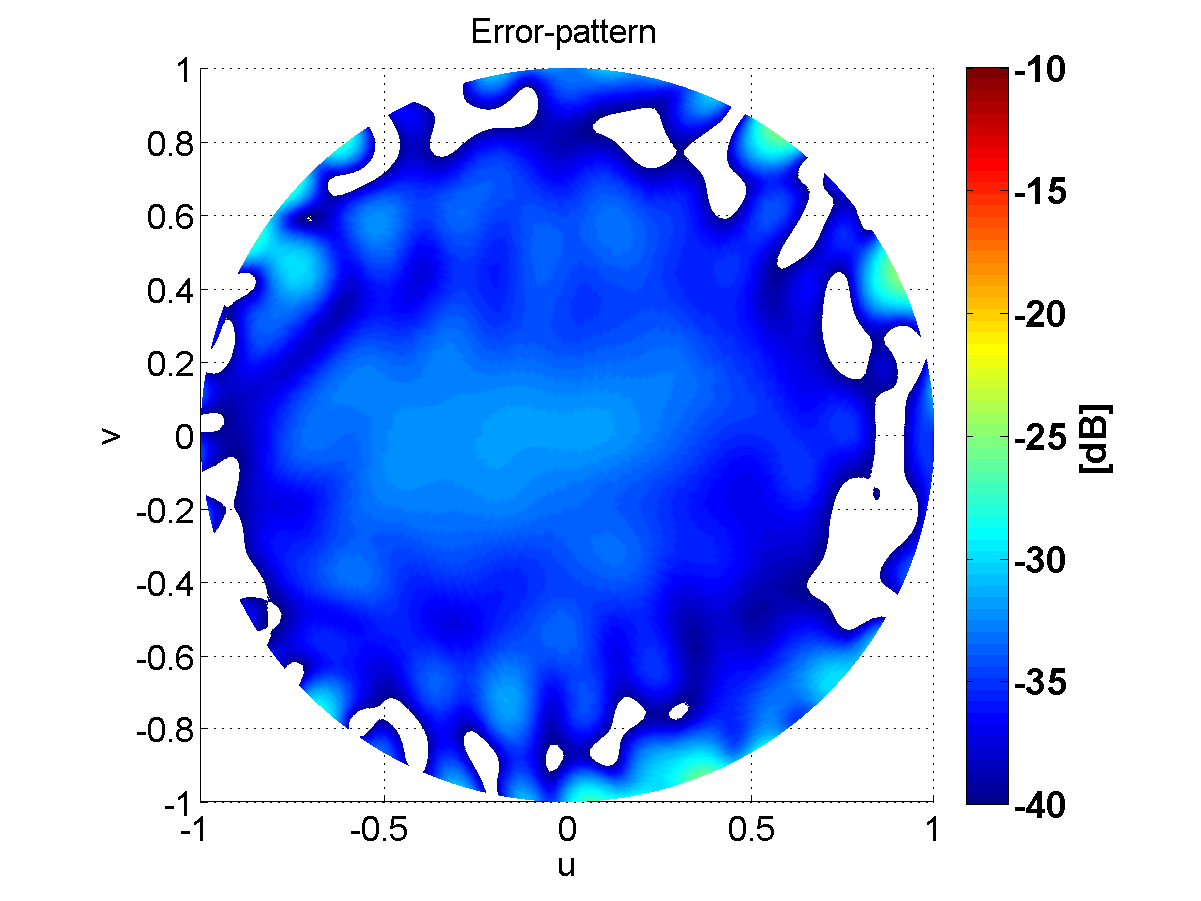}}

\subfigure[110\,MHz.]{\includegraphics[scale=0.235,clip,trim={1.25cm 0cm 1cm 0cm}]{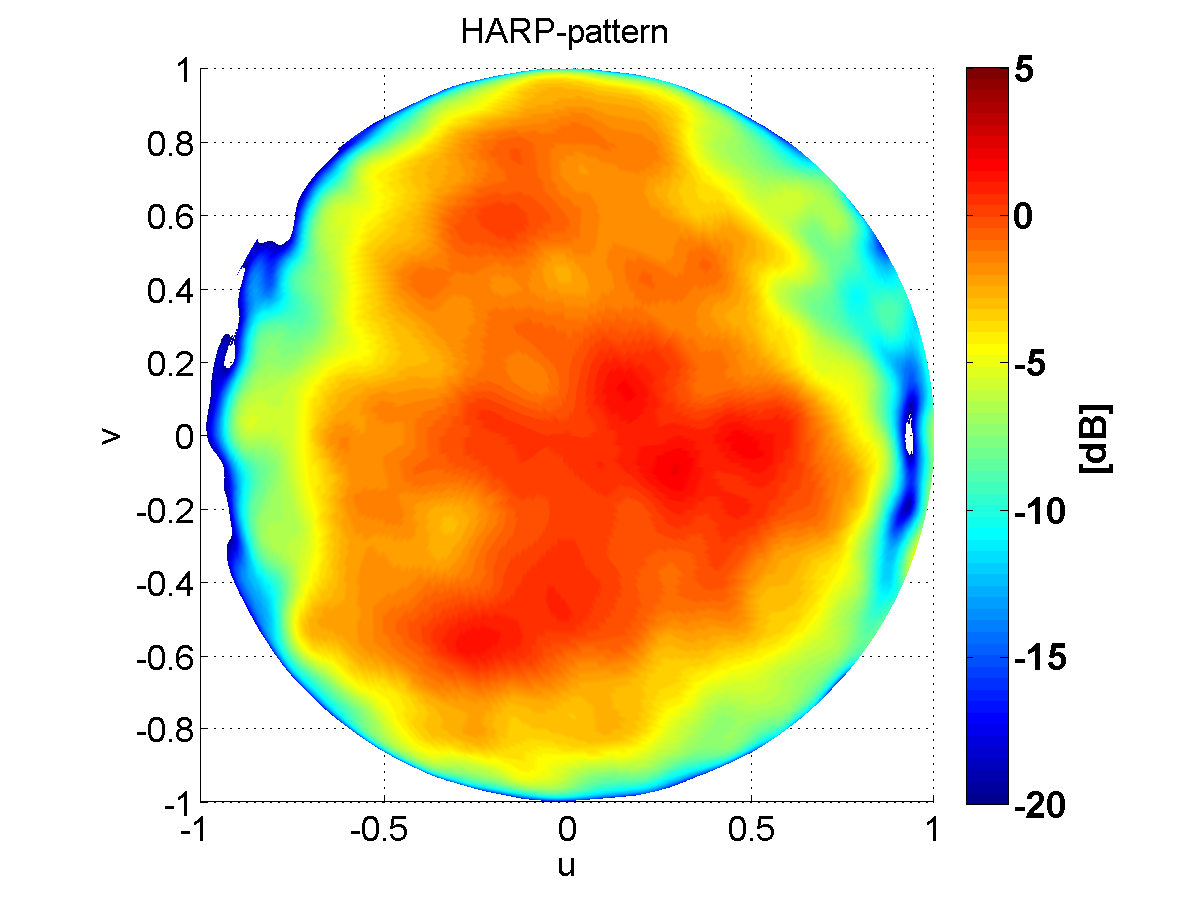}}
\subfigure[Error Pattern 110\,MHz]{\includegraphics[scale=0.235,clip,trim={1.25cm 0cm 1cm 0cm}]{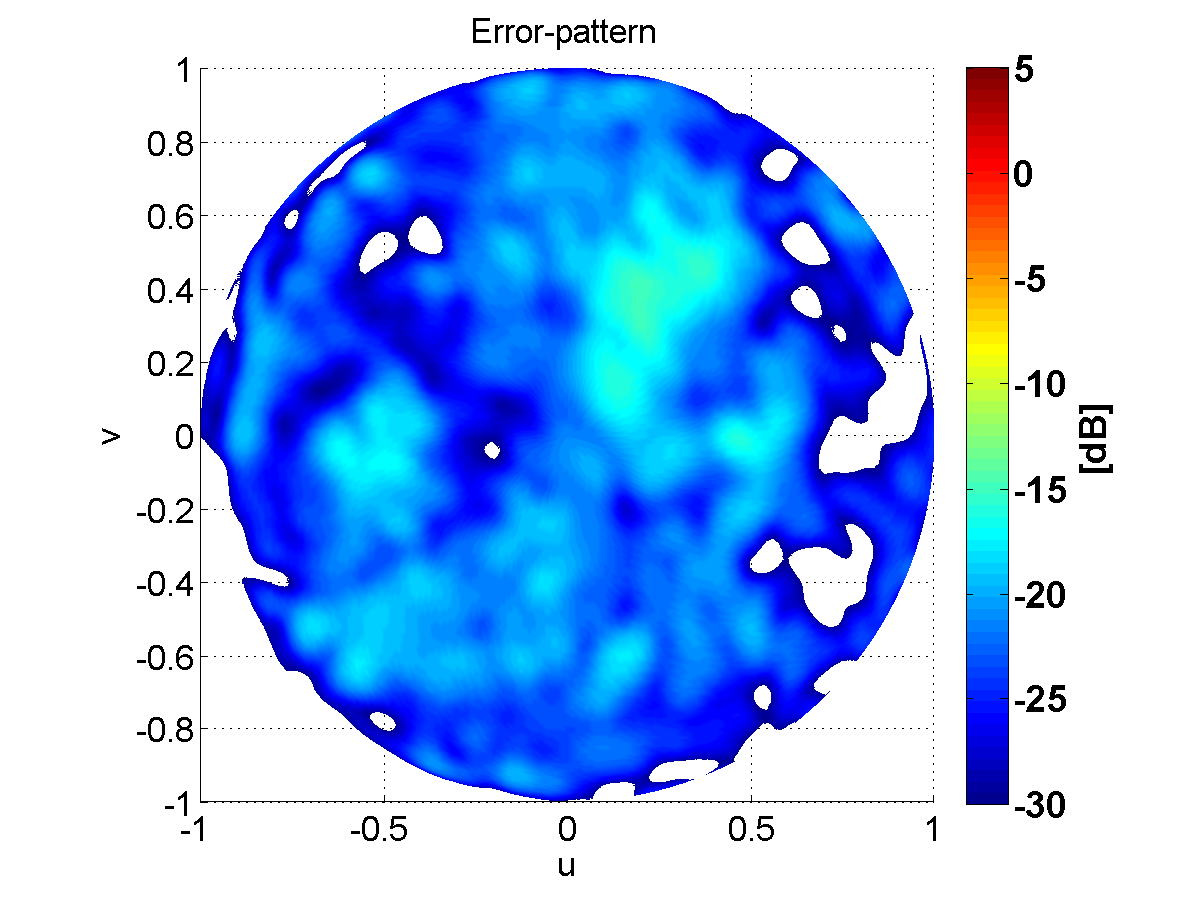}}

\subfigure[200\,MHz.]{\includegraphics[scale=0.235,clip,trim={1.25cm 0cm 1cm 0cm}]{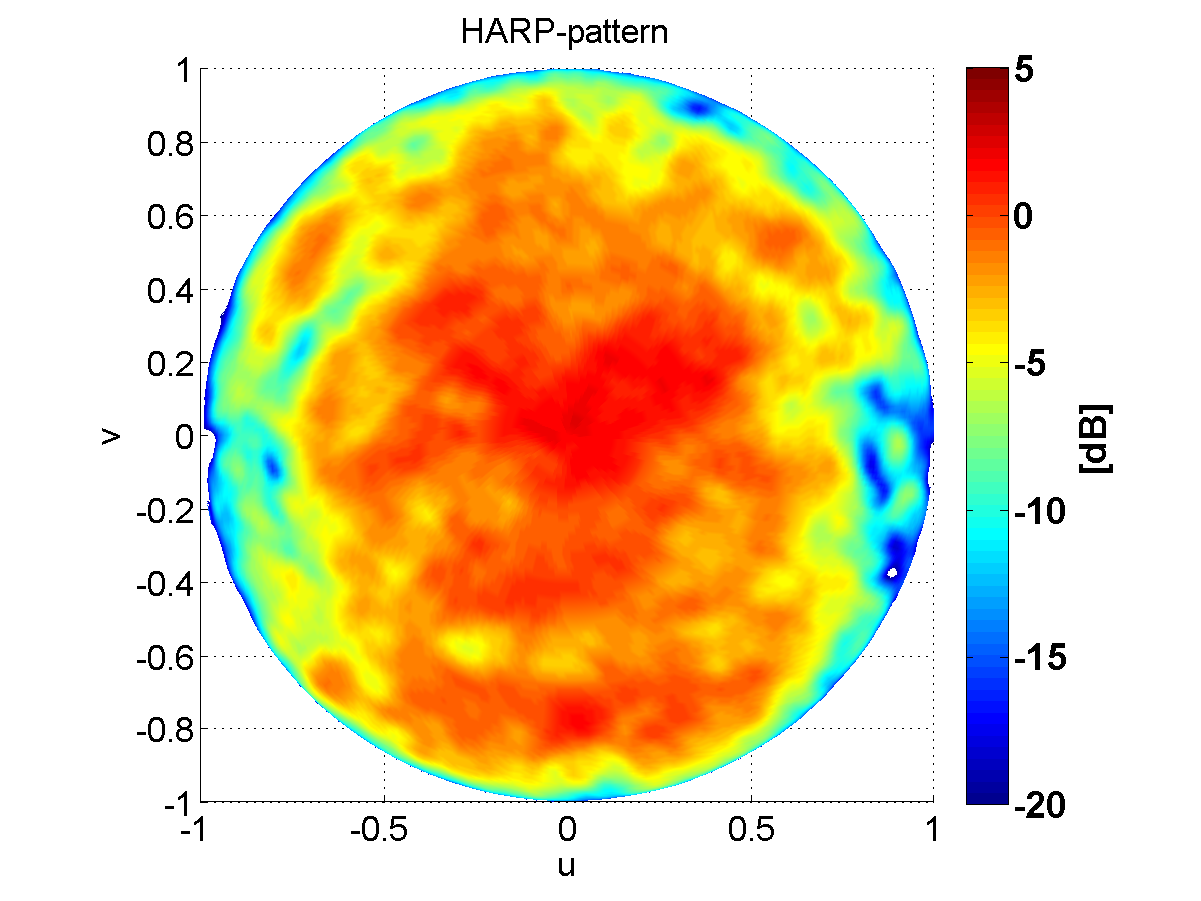}}
\subfigure[Error Pattern 200\,MHz]{\includegraphics[scale=0.235,clip,trim={1.25cm 0cm 1cm 0cm}]{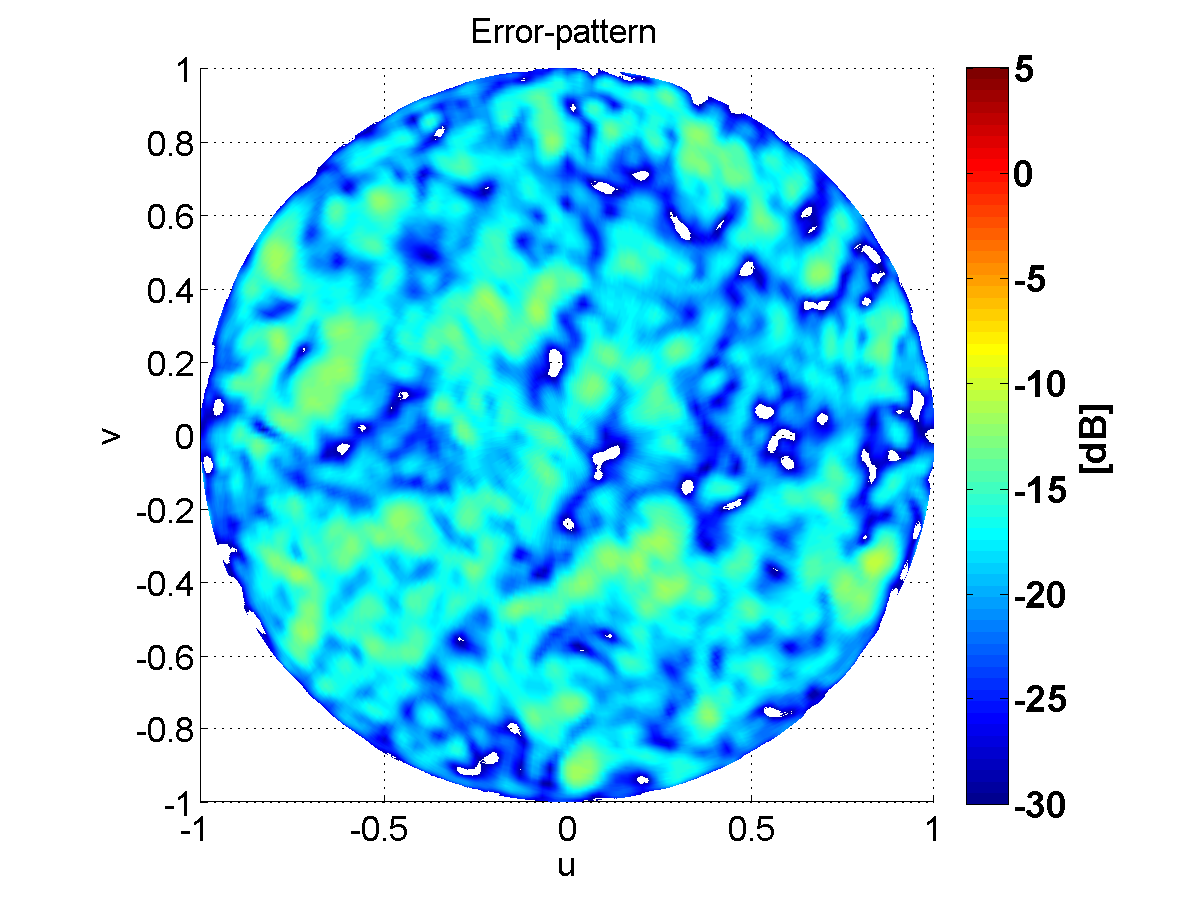}}

\subfigure[350\,MHz.]{\includegraphics[scale=0.235,clip,trim={1.25cm 0cm 1cm 0cm}]{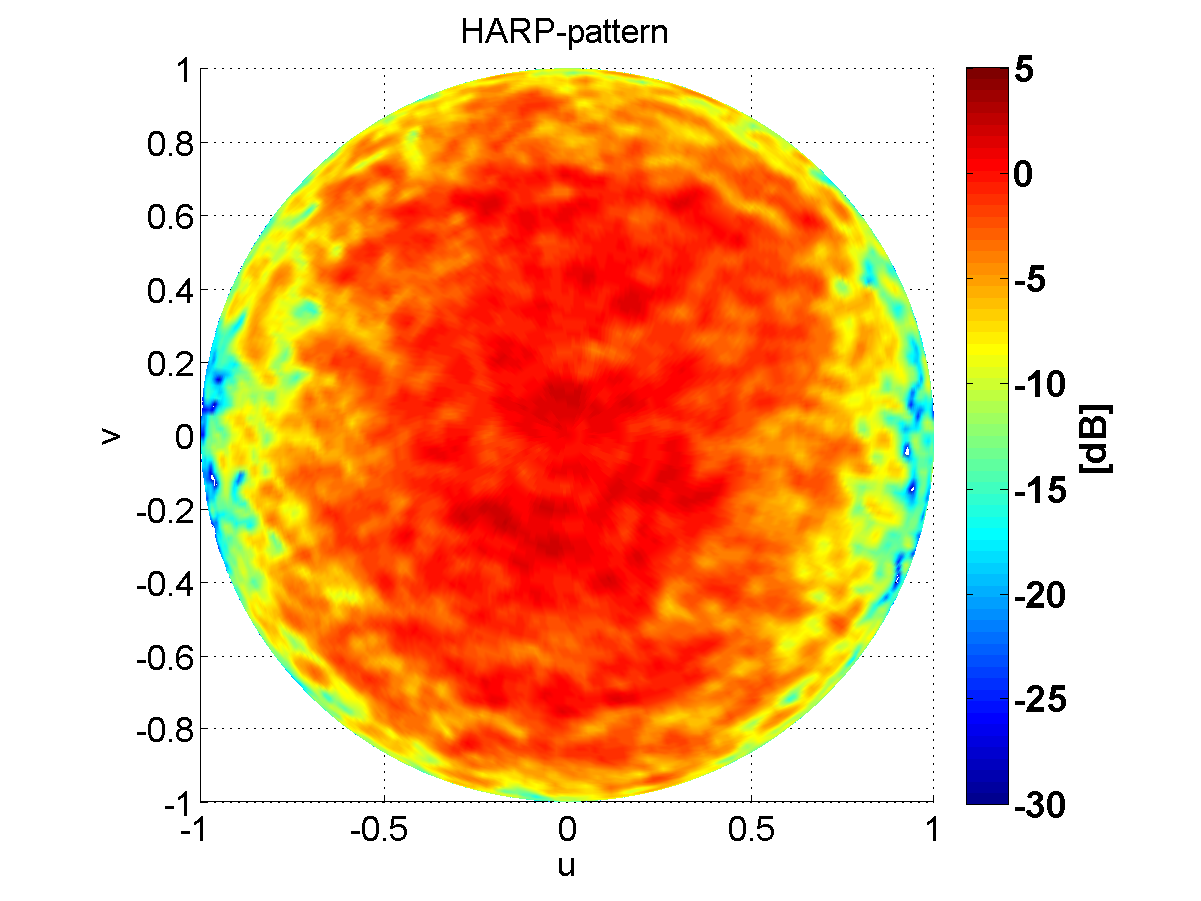}}
\subfigure[Error Pattern 350\,MHz]{\includegraphics[scale=0.235,clip,trim={1.25cm 0cm 1cm 0cm}]{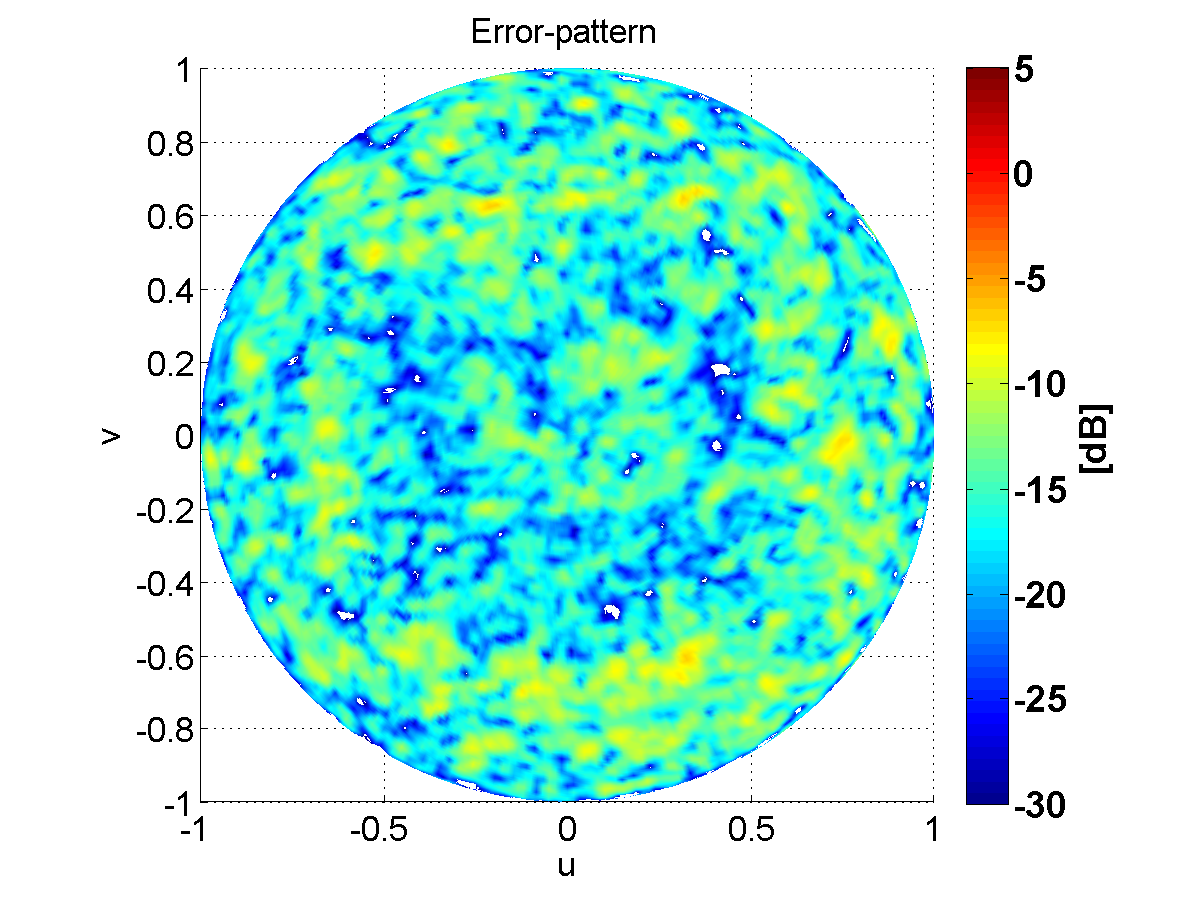} \label{fig:sub_eep_h}}

\caption{EEP obtained by HARP and the difference w.r.t. WIPL-D of a random element in the SKA station (red dot in Fig.~\ref{fig:array256}).}
\label{fig:eep}

\end{figure}

\subsection{Embedded Element Patterns}

In applications such as the SKA telescope, the EEPs are of primary interest as they allow one to form different beams to get information of signal from sky. An accurate and effective calculation of all EEPs is then required, while taking  into account the mutual coupling. In this section, the quality of EEPs obtained using HARP is evaluated. A small testbed of 16 SKALA elements is first analyzed, where the EEPs of all elements are calculated at all frequencies with 1\,MHz frequency steps. Then, the SKA station in Fig.~\ref{fig:array256} is simulated at four different frequencies, i.e. 50\,MHz, 110\,MHz, 200\,MHz and 350\,MHz, using HARP and WIPL-D, and the calculated EEPs are compared.

\subsubsection{EEPs of A Small Array}
The performance of HARP has been tested against the brute force MoM solution of a small array of 16 SKALA elements for the frequency band (50 -- 450\,MHz). 
The array corresponds to a testbed for SKA1-Low technology and is situated at the Mullard Radio Astronomy Observatory (MRAO) near Cambridge, UK~\cite{Eloy15}.
This array with only 16 elements is in a pseudo-random configuration, such as that of SKA stations, and has the same expected element density, with an average inter-element spacing of half wavelength at 77\,MHz.
The error between the patterns is calculated as:
\begin{equation}
\epsilon = \sqrt{\left|E_\theta - E_\theta^{ref}\right|^2 + \left|E_\phi - E_\phi^{ref}\right|^2}
\label{eq:eep_error}
\end{equation} %
where $E$ and $E^{ref}$ being the EEP calculated by HARP and brute-force MoM with the electric fields being normalized by the maximum of the isolated antenna pattern.
This error is calculated in all directions for all involved antennas.
The condensed result is shown in Fig. \ref{fig:eep_error} by calculating the percentiles of the error and plotting them over frequency.
The HARP method shows very good accordance to the brute-force MoM results over the whole frequency range. 
The error increases  at the highest end of the band, but it is still below -30 dB.

\begin{figure}[!htb]
\centering
\begin{tikzpicture}
\node at(0,0){
\includegraphics[scale=0.65,clip,trim={0.5cm 0cm 0.5cm 0cm}]{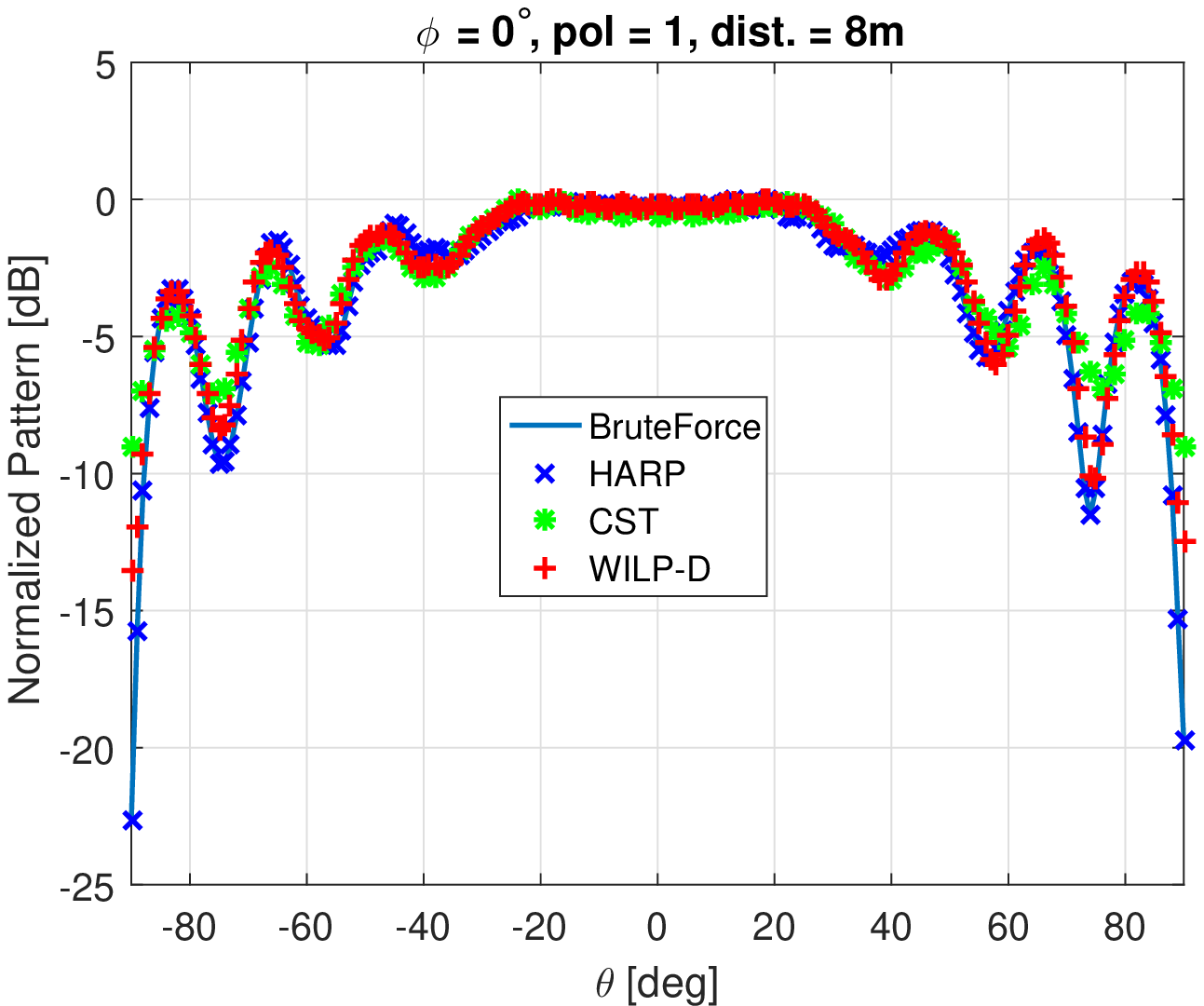}};
\node at(0.09,-0.7){\includegraphics[scale=0.425,clip,trim={.65cm 0cm 0.5cm 1cm},height=3cm,width = 5.65cm]{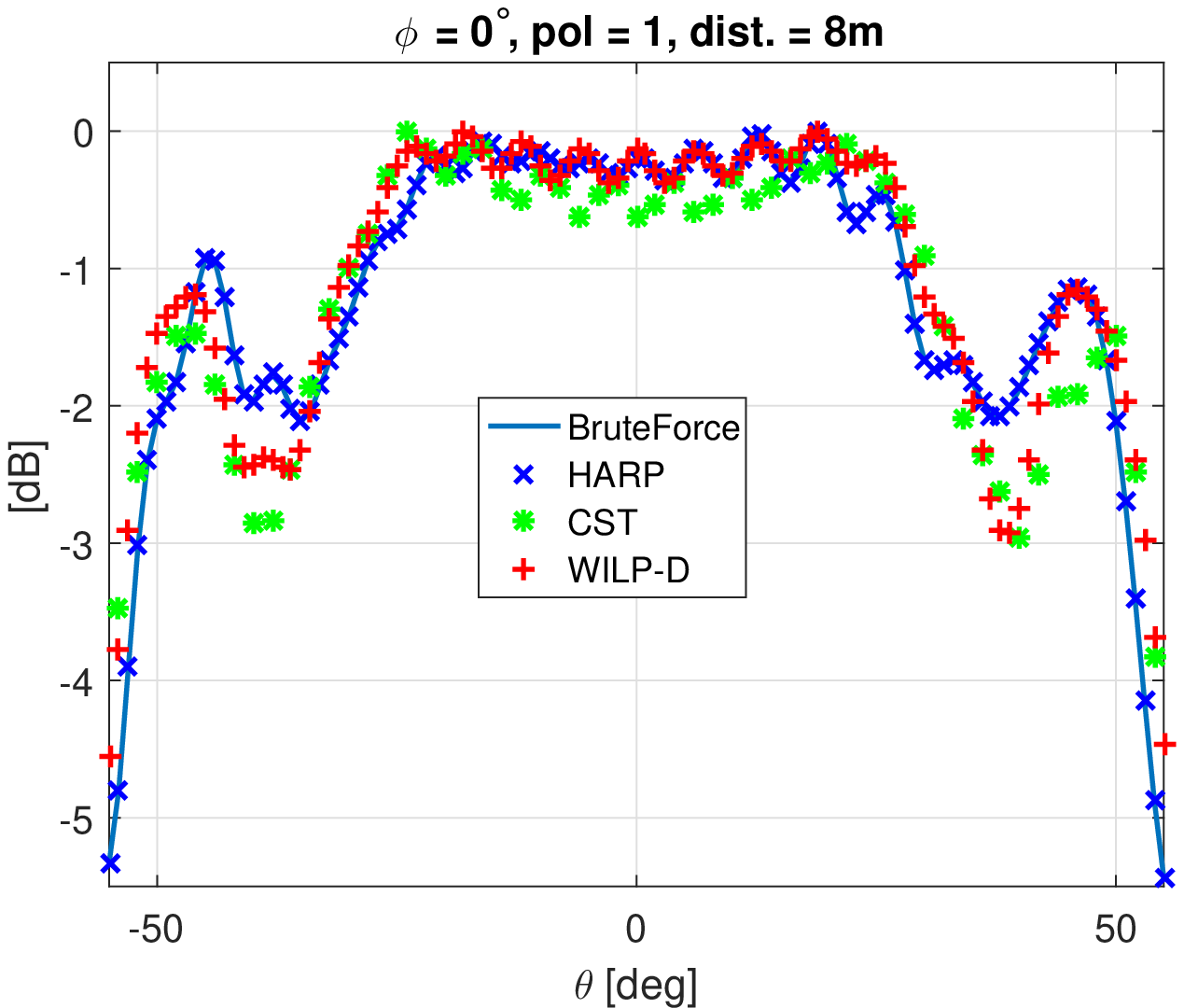}};
\end{tikzpicture}
\caption{$\phi = 0^\circ$ cut of one EEP in an array of 2 SKALA at a distance of 8m for the first polarization at 350\,MHz. (Inset: zoom in of the main beam)}
\label{fig:array2}
\end{figure}

\begin{figure}[!htb]\centering

\subfigure[E-plane: $\phi = 0^\circ$, 50\,MHz]{\includegraphics[scale=0.315,clip,trim={0.25cm 0cm 1cm 0cm}]{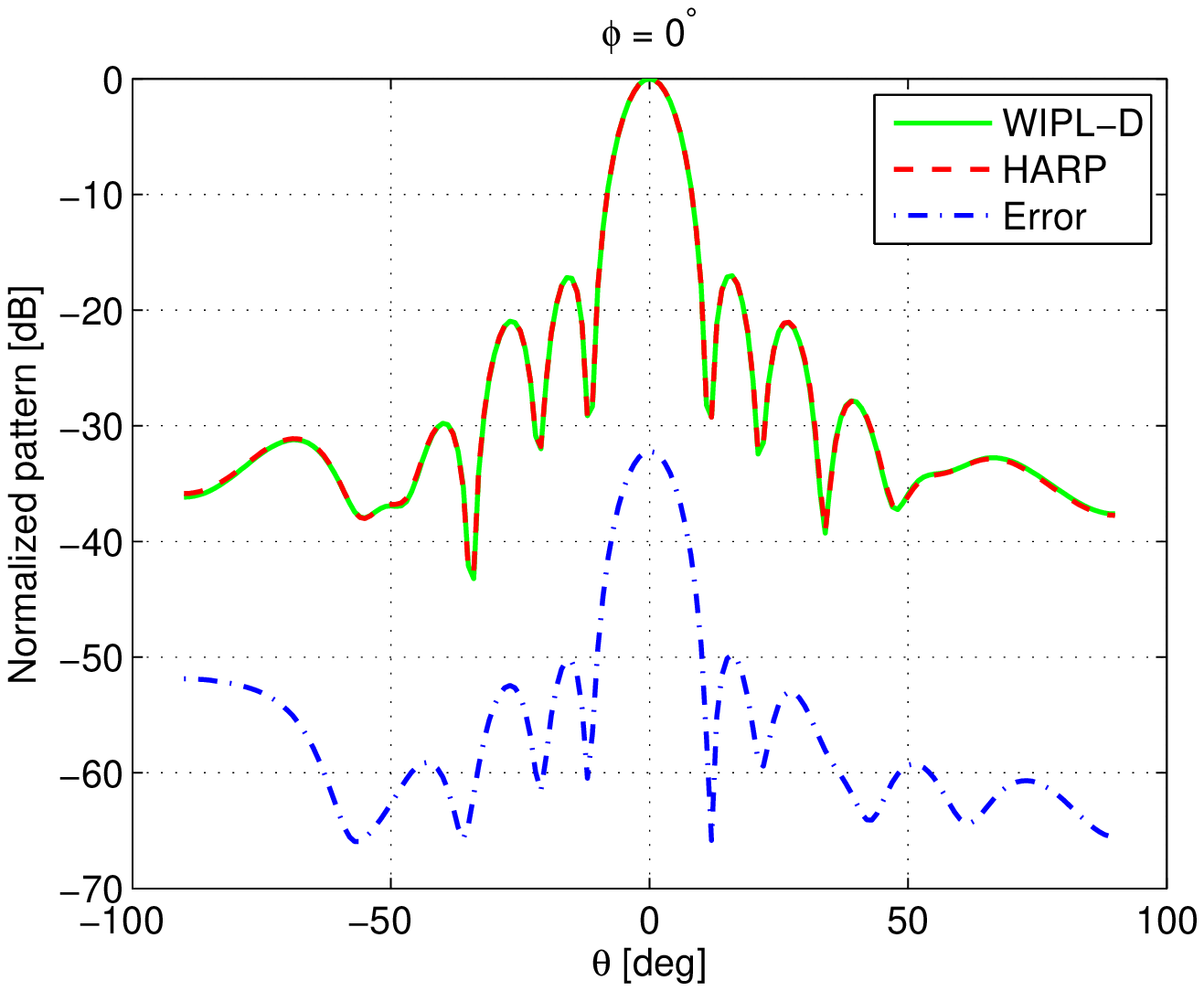}}
\subfigure[H-plane: $\phi = 90^\circ$, 50\,MHz]{\includegraphics[scale=0.315,clip,trim={0.25cm 0cm 1cm 0cm}]{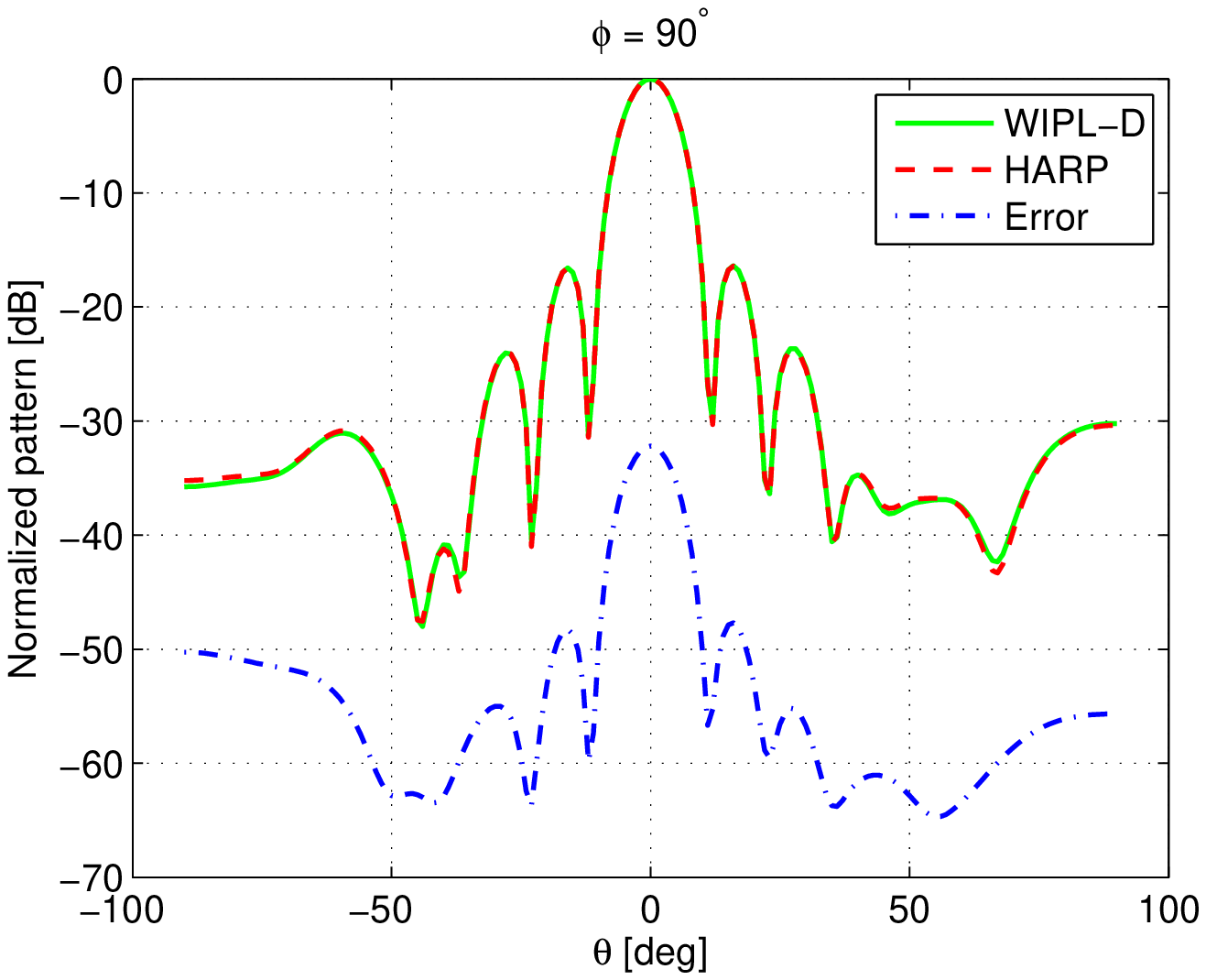}}

\subfigure[E-plane: $\phi = 0^\circ$, 110\,MHz]{\includegraphics[scale=0.315,clip,trim={0.25cm 0cm 1cm 0cm}]{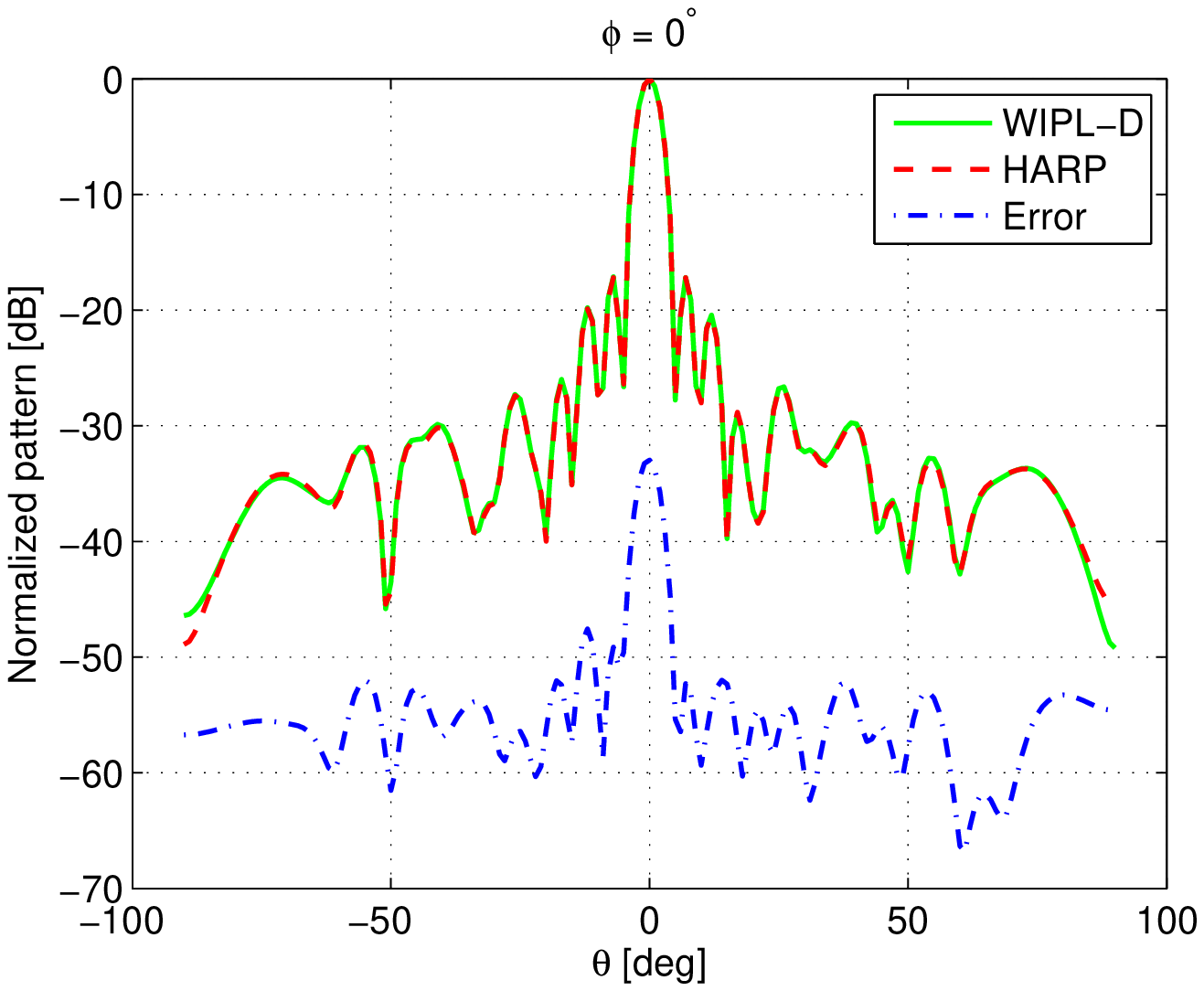}}
\subfigure[H-plane: $\phi = 90^\circ$, 110\,MHz]{\includegraphics[scale=0.315,clip,trim={0.25cm 0cm 1cm 0cm}]{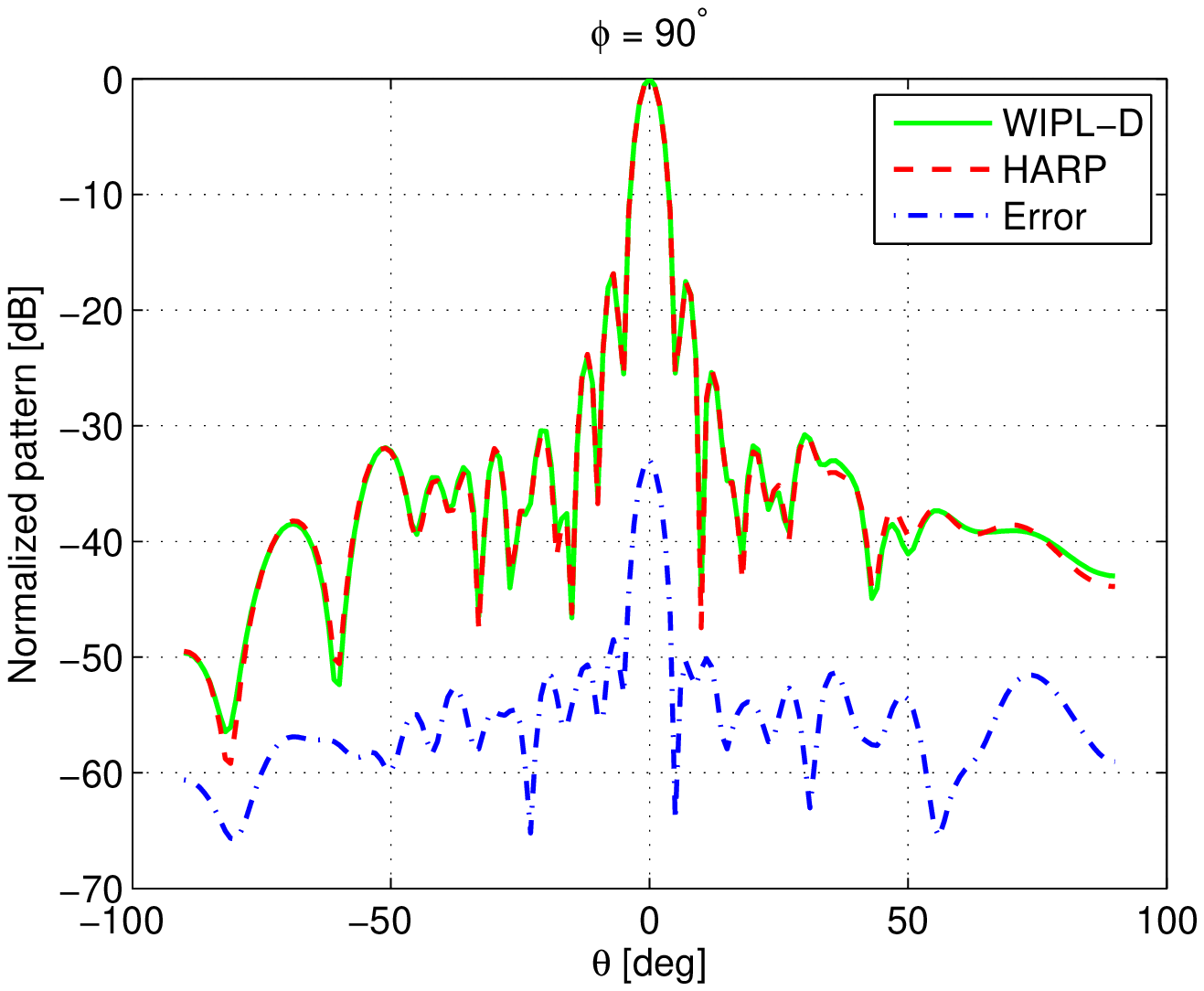}}

\subfigure[E-plane: $\phi = 0^\circ$, 200\,MHz]{\includegraphics[scale=0.315,clip,trim={0.25cm 0cm 1cm 0cm}]{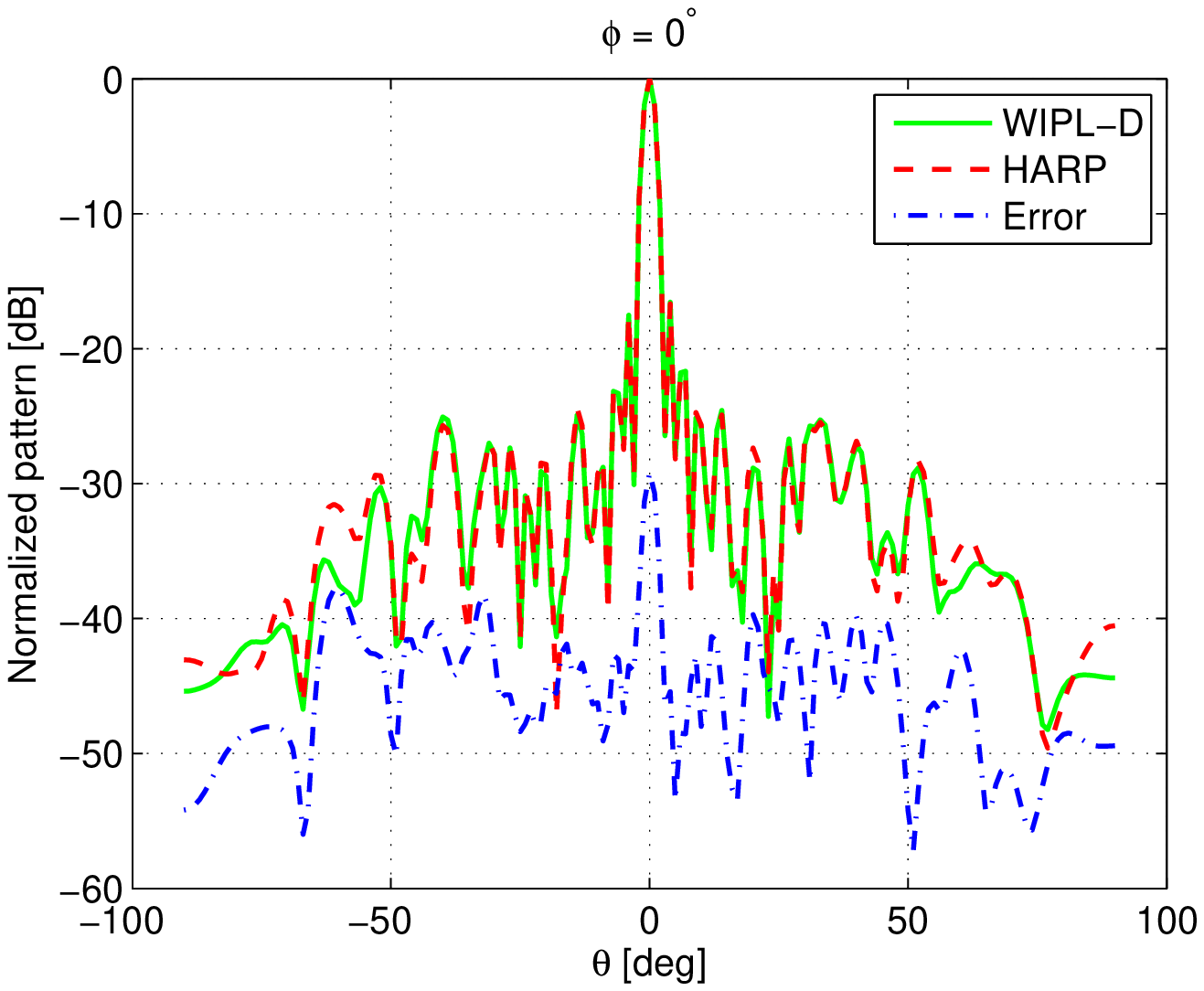}}
\subfigure[H-plane: $\phi = 90^\circ$, 200\,MHz]{\includegraphics[scale=0.315,clip,trim={0.25cm 0cm 1cm 0cm}]{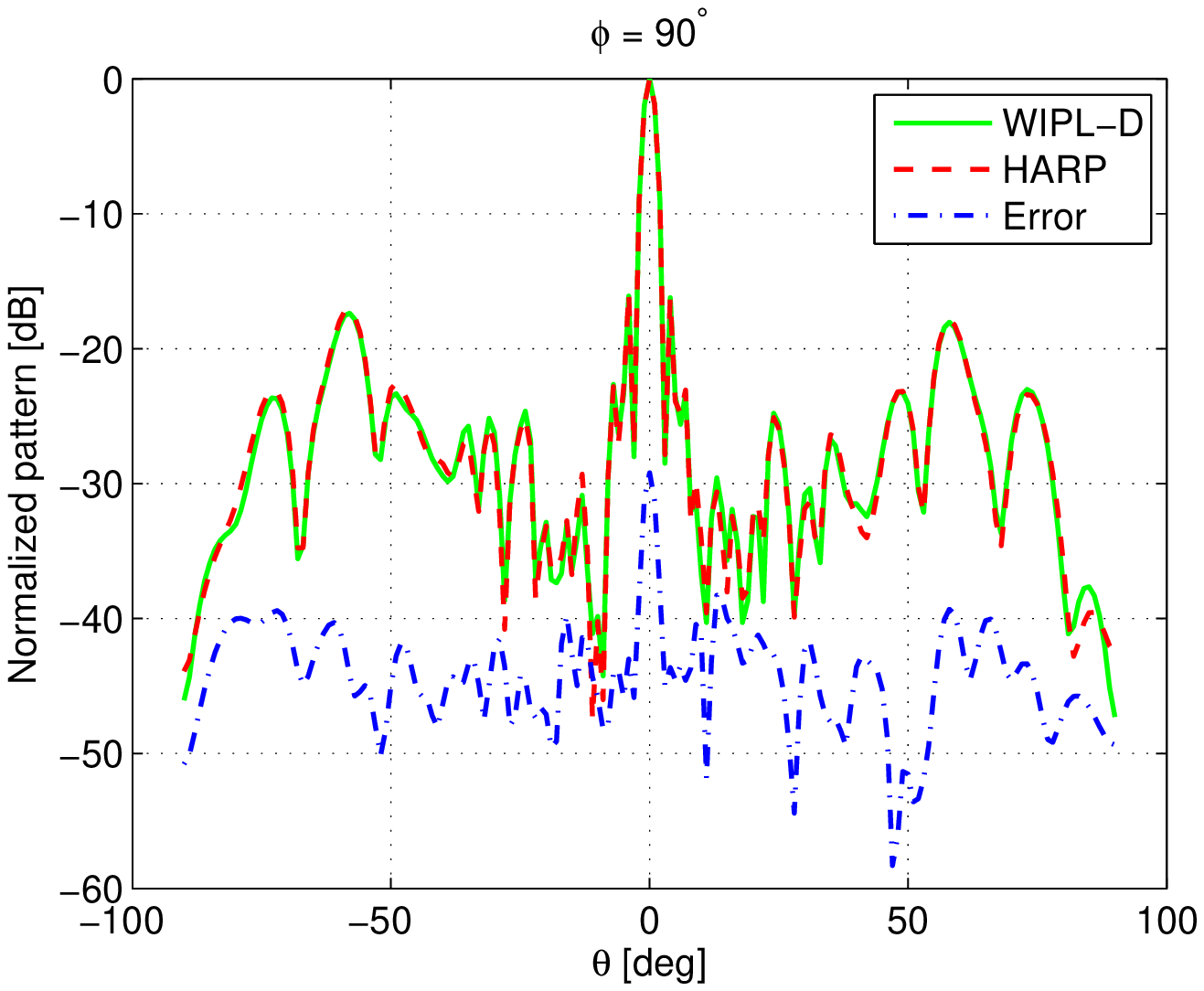}}

\subfigure[E-plane: $\phi = 0^\circ$, 350\,MHz]{\includegraphics[scale=0.315,clip,trim={0.25cm 0cm 1cm 0cm}]{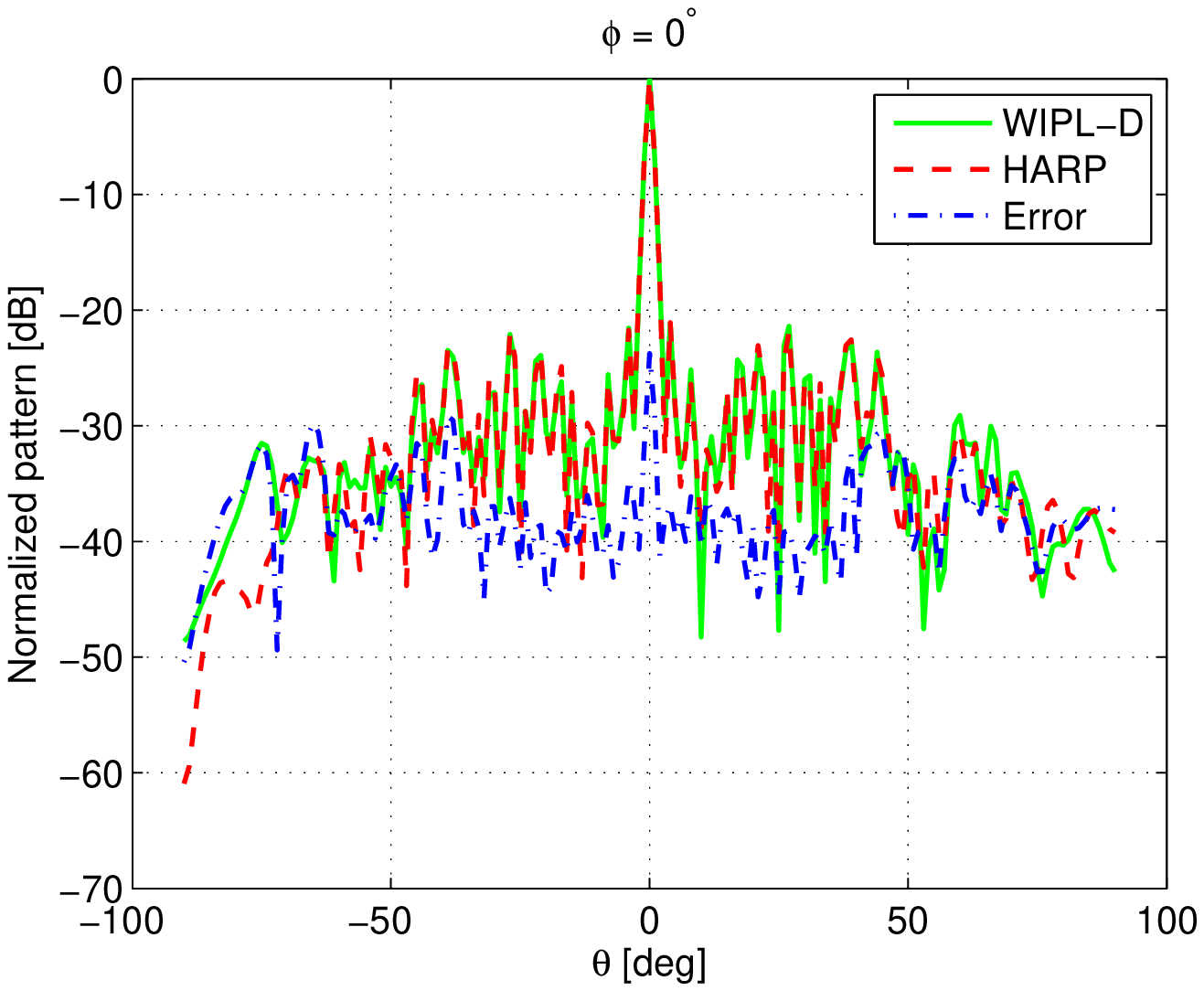}}
\subfigure[H-plane: $\phi = 90^\circ$, 350\,MHz]{\includegraphics[scale=0.315,clip,trim={0.25cm 0cm 1cm 0cm}]{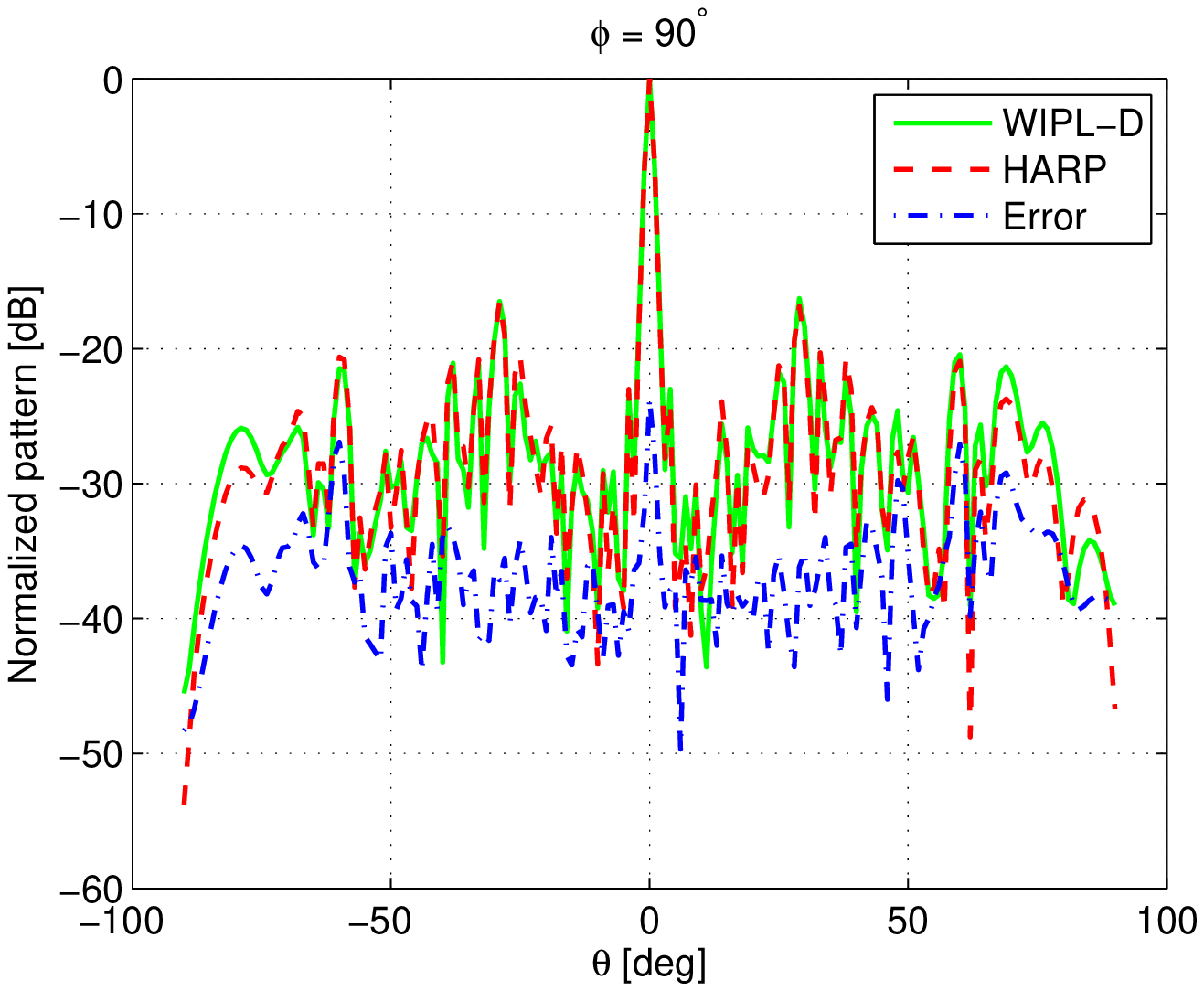} \label{fig:sub1}}

\caption{Normalized radiation pattern of a SKA station in two principle planes at different frequencies.}
\label{fig:radStation}

\end{figure}

\subsubsection{EEPs of SKA station}
The performance of HARP on the SKA station is now validated with patterns simulated using  the WIPL-D software~\cite{wipld}.
The EEP of an arbitrary element (red dot in Fig.~\ref{fig:array256}) obtained using HARP is taken as an example. Fig.~\ref{fig:eep} shows the EEPs at four different frequencies and their differences w.r.t. WIPL-D, where the error patterns are computed using (\ref{eq:eep_error}) with $E$ and $E^{ref}$ being patterns obtained using HARP and WIPL-D, respectively.
An excellent agreement is shown at low-frequencies, where the difference level is found below -30\,dB at 50\,MHz. The variation becomes more visible as the frequency increases. 
A detailed study of the performance of different software at 350\,MHz is carried out on an array of 2 antennas positioned along the $x-$axis with varying distance. The EEPs obtained using CST, WIPL-D, brute-force MoM and HARP are plotted together in Fig.~\ref{fig:array2} for a separation equal to 8\,m. For this 2 element array, the EEPs are similar to the isolated pattern, and mutual coupling clearly appears as ripples in the main beam. The results show that while the mutual coupling is accounted for in all software packages, a small difference is observed in the EEPs. 
The reason is attributed to the difference in the geometrical modeling of the SKALA, i.e. the use of different meshes in each software, which leads to differences in the isolated pattern. At the highest frequency, a small difference in antenna modeling will result in a significant variation in the radiation patterns. This explains the discrepancies in the EEPs visible in Fig.~\ref{fig:sub_eep_h}, as well as the self-impedance behavior in Fig.~\ref{fig:Z11}.
A stronger indicator that the discrepancies mainly come from the mesh is that the errors in the isolated pattern are almost as large as in the EEPs.
 Moreover, it is important to note that, in Fig.~\ref{fig:array2}, patterns calculated using brute-force MoM and HARP overlap very well, which confirms the accuracy of HARP method.

\subsection{Array Patterns}

\begin{figure}[!htb]
\centering
\includegraphics[scale=0.65,clip,trim={0.5cm 0cm 0.5cm 0cm}]{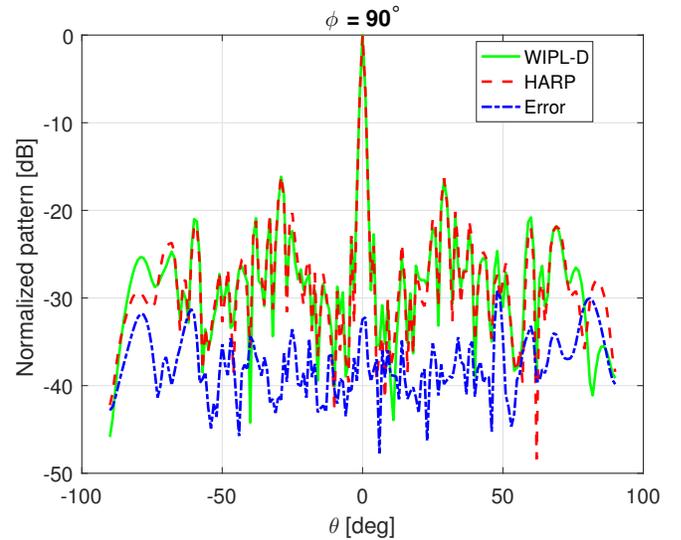}
\caption{Normalized radiation pattern of a SKA station at 350\,MHz after the difference between isolated pattern is added to all EEPs; in H-plane ($\phi = 90^\circ$) cut.}
\label{fig:ska350_corrected}
\end{figure}

In this section, the EEPs calculated using HARP and WIPL-D software are exploited to form the array pattern. A simple beamforming method is implemented with uniform excitation and scanning at broadside. Fig.~\ref{fig:radStation} shows the radiation pattern of the SKA station at different frequencies in two principal planes calculated using HARP and WIPL-D. The two methods show a great agreement across the frequency band, as the difference in patterns, evaluated by (\ref{eq:eep_error}), is only shown at a -30\,dB level, except for the highest frequency, i.e. 350\,MHz, where the error in the main beam is near -24\,dB. 
As explained above, the visible difference in the patterns at higher frequency is due to the difference in modeling of the SKALA in the two software.
To demonstrate the impact of the isolated element modeling, we show that it  can be partially undone by properly adding the difference of isolated patterns between HARP and WIPL-D into the EEPs, as shown in Fig.~\ref{fig:ska350_corrected}. A better agreement is clearly seen w.r.t. Fig.~\ref{fig:sub1}, especially near the main beam.  This exercise illustrates that the difference between WIPL-D and HARP at highest frequency is mainly due to the modeling of the SKA antennas in these two software packages. Expectedly, this fine modeling difference becomes more visible at high-frequency.

\begin{table}[!t]
\renewcommand{\arraystretch}{1.3}
\caption{Computational Time Obtaining Embedded Element Patterns of a SKA1-Low station at 110\,MHz}
\label{tab:ComputTime}
\centering
\begin{tabular}{c c c c }
\hline
 & \textbf{CST} & \textbf{WIPL-D} & \textbf{HARP} \\
\hline
\hline
Simulation Time & 96 hours  & 97 hours & 0.5 min.\\
\hline
\end{tabular}
\end{table}

Finally, the performance of WIPL-D, CST and HARP for the analysis of a SKA station at 110\,MHz is detailed in Table~\ref{tab:ComputTime}. The CST simulations were carried out on a server with two processors and two cores each with 384\,GB of RAM at the University of Cambridge, UK, whereas the WIPL-D ones were performed on a Windows Server 2008 R2 Enterprise with  an Intel Xeon X7550 processor and 125\,GB RAM at ASTRON, The Netherlands. All the simulations for HARP were carried out on a desktop computer with 16\,GB RAM and Processor Intel Core $i7 - 3770$ CPU 3.4\,GHz at UCL. For HARP, the accounted time stands for matrix filling, solving the system of equations and calculation of all EEPs once the HARP model is built, as reported in Table~\ref{tab:harpTime}. 
A great time saving factor is achieved using HARP, while providing highly accurate results, as demonstrated in Fig.~\ref{fig:radStation}.
The establishment of the HARP model takes about 3 hours per frequency. It is important to recall that this preparation phase takes place only once and for all, regardless of the array configuration, and can be re-used to analyze any array made of the same element. This feature is very beneficial for simulating SKA arrays, where hundreds of stations are planned, as well as to analyze extremely large arrays, as will be discussed in the next section.

\begin{table}[!ht]
\renewcommand{\arraystretch}{1.3}
\caption{Analysis Time for a SKA station using HARP at 110\,MHz.}
\label{tab:harpTime}
\centering
\begin{tabular}{|c||c|}
\hline
\textbf{Operations} & \textbf{ Time (in seconds)}\\
\hline
Matrix Filling & 10 \\
\hline
Solving MoM equations & 12.6\\
\hline
Calculation EEPs & 6 \\
\hline
\textbf{Total} & \textbf{28.6} \\
\hline
\end{tabular}
\end{table}

\begin{figure}[!htb]\centering
\includegraphics[scale=0.8,clip,trim={1.5cm 0cm 2cm 0cm}]{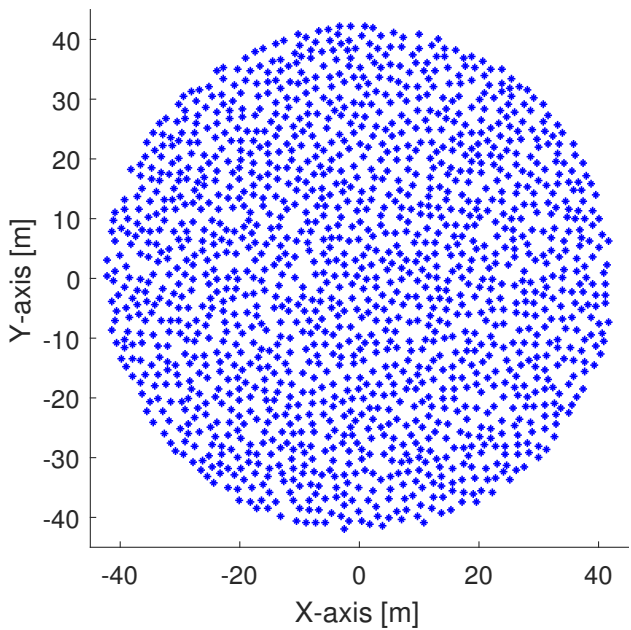}
\caption{Layout of 1500 SKALA element array with minimum distance of 1.35\,m.}
\label{fig:array1500}
\end{figure}

\begin{figure*}[!htb]
\subfigure[E-plane: $\phi= 0^\circ$.]{\includegraphics[scale=0.485,clip,trim={0.5cm 0cm 0.5cm 0cm}]{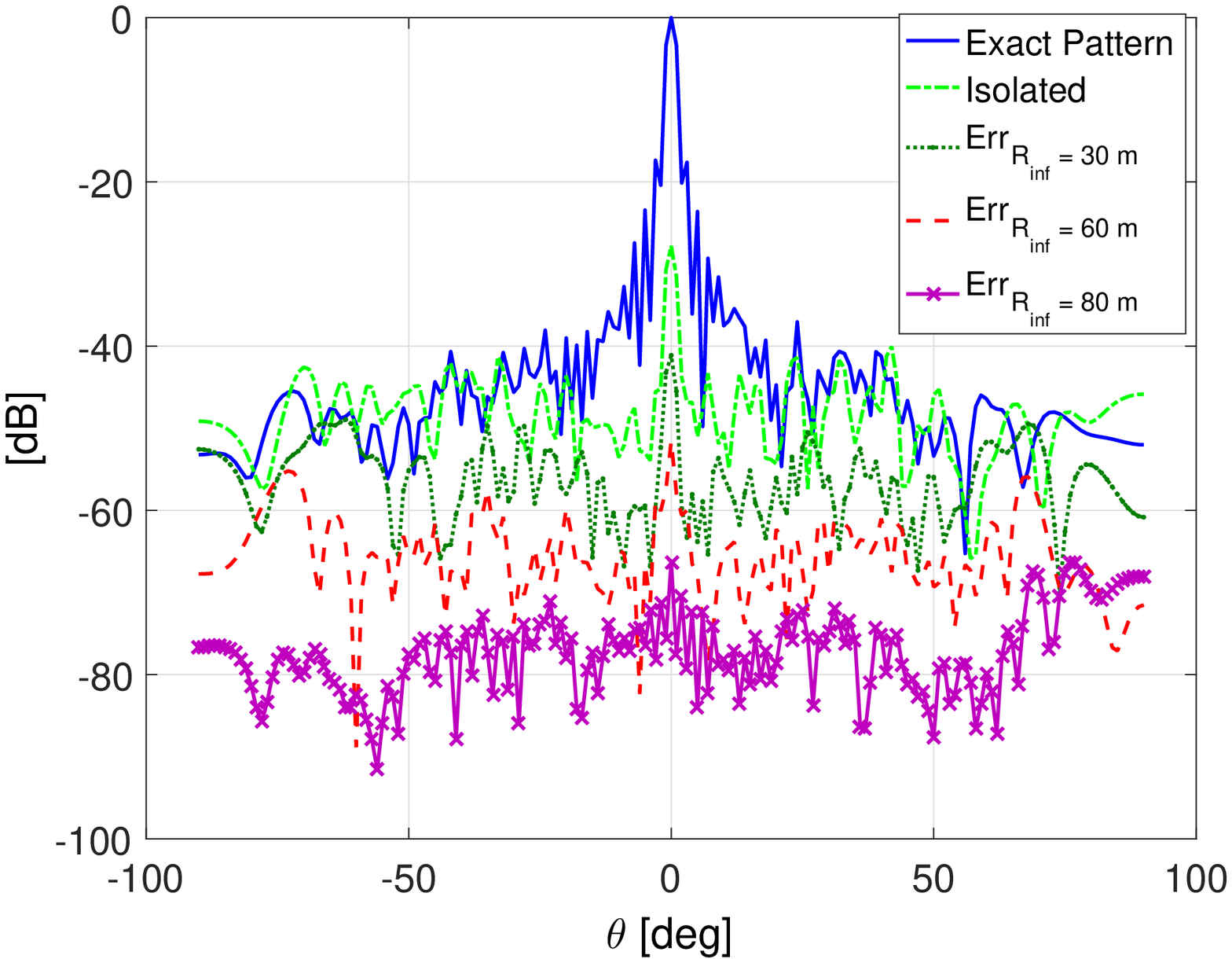}}
\subfigure[H-plane: $\phi= 90^\circ$.]{\includegraphics[scale=0.485,clip,trim={0.5cm 0cm 0.5cm 0cm}]{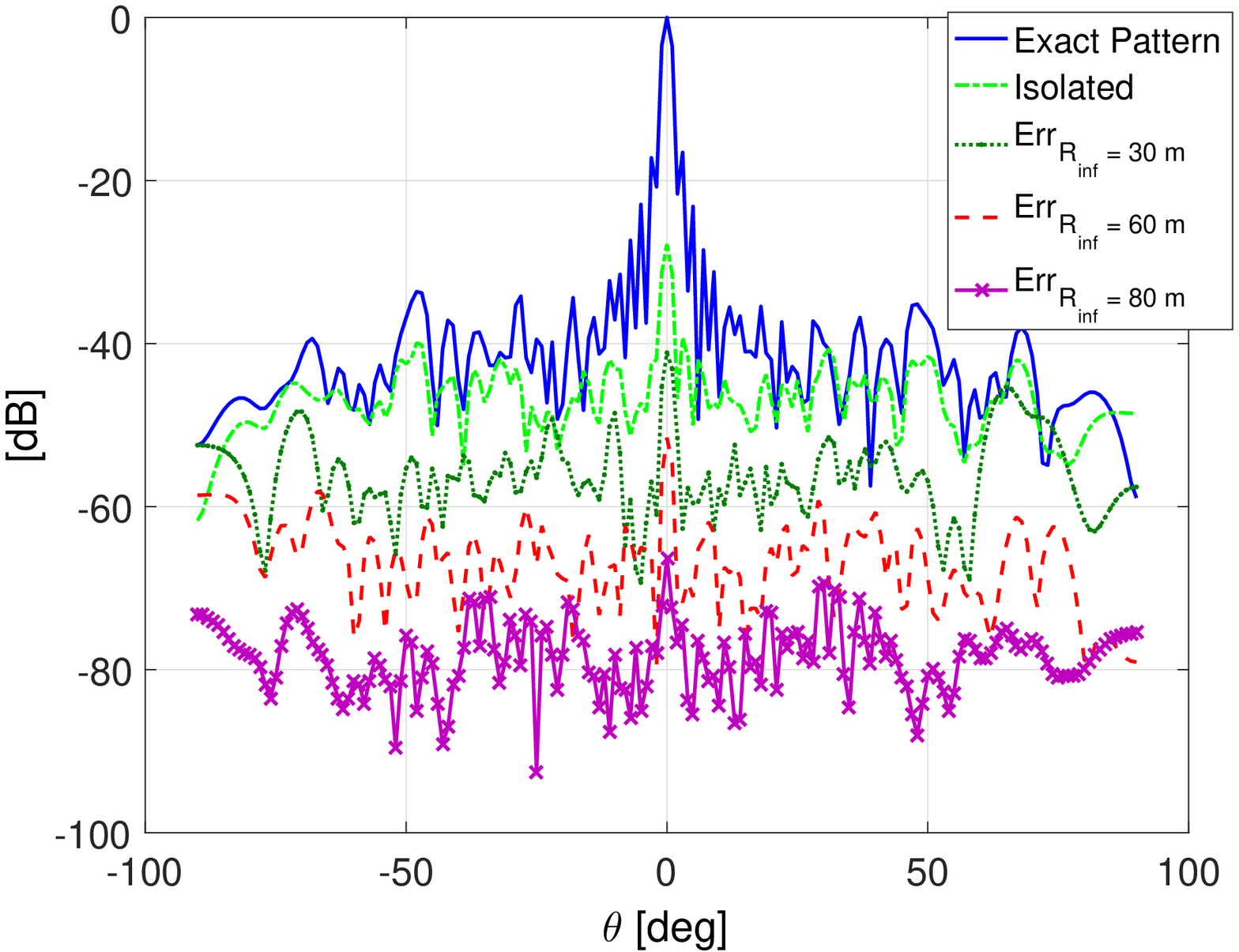}}

\caption{Normalized radiation pattern of the array in Fig.~\ref{fig:array1500} at 110\,MHz and errors for different RIs in two principal planes using the sparse matrix approach. (a) E-plane, and (b) H-plane.}
\label{fig:array1500RI}
\end{figure*}

\begin{figure}[!htb]\centering
\includegraphics[scale=0.475,clip,trim={0cm 0cm 0cm 0cm}]{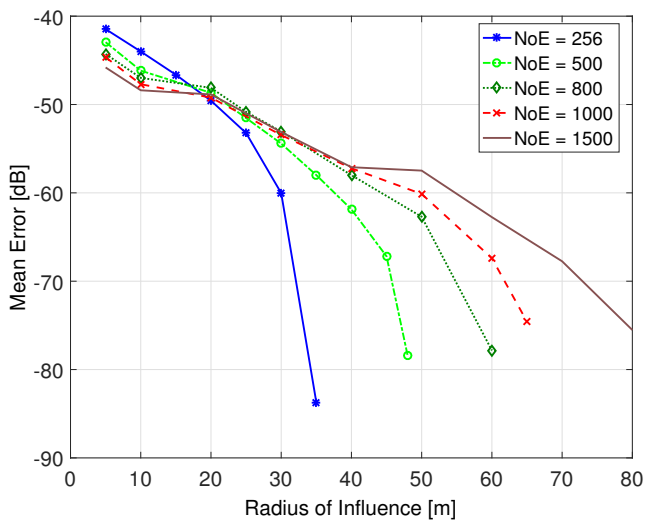}
\caption{Mean errors vs. RI for different array sizes using the sparse matrix approach at 110\,MHz.}
\label{fig:RIconvergence}
\end{figure}

\begin{figure*}[!htb]
\subfigure[E-plane: $\phi= 0^\circ$.]{\includegraphics[scale=0.475,clip,trim={0.5cm 0cm 0.5cm 0cm}]{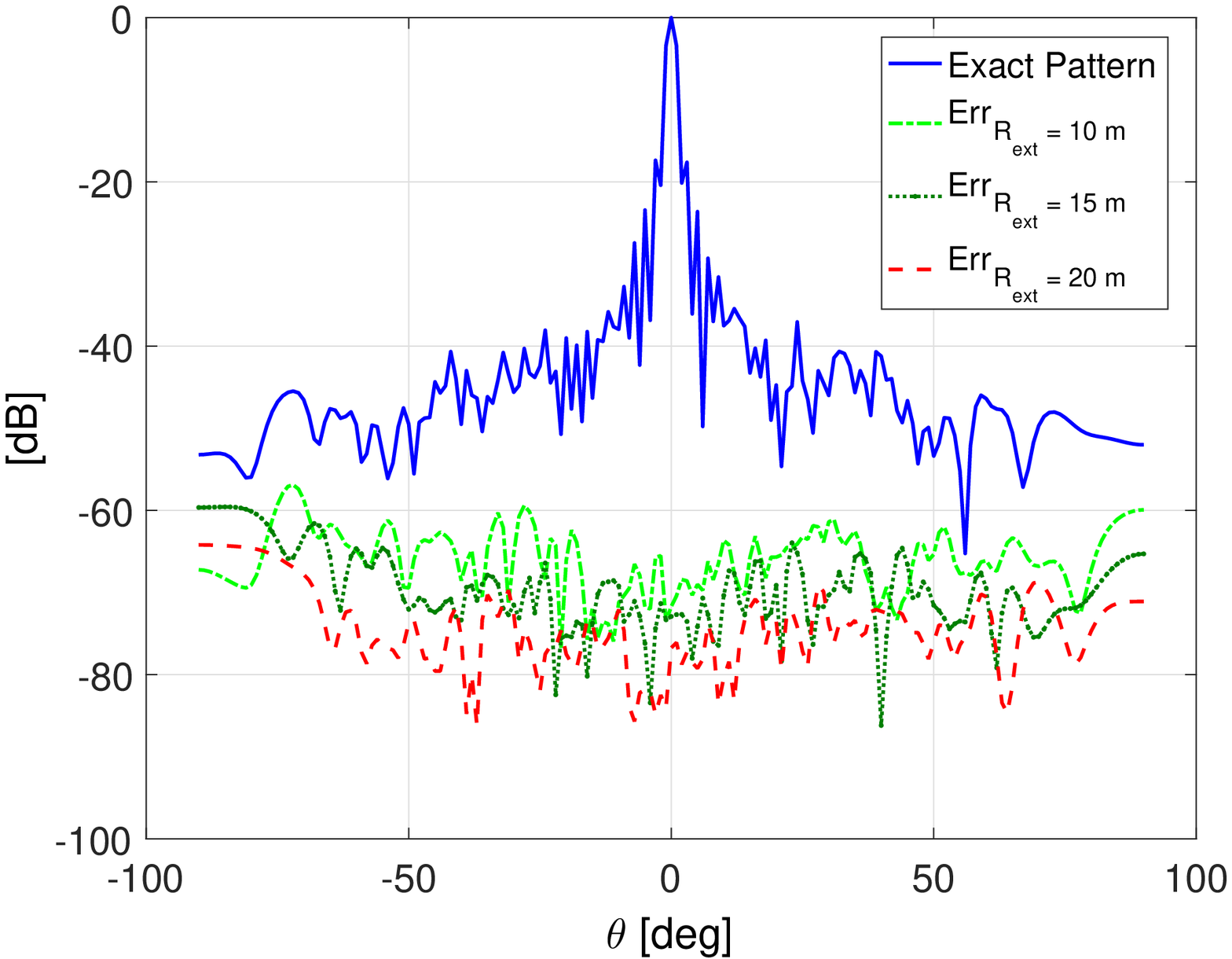}}
\subfigure[H-plane: $\phi= 90^\circ$.]{\includegraphics[scale=0.475,clip,trim={0.5cm 0cm 0.5cm 0cm}]{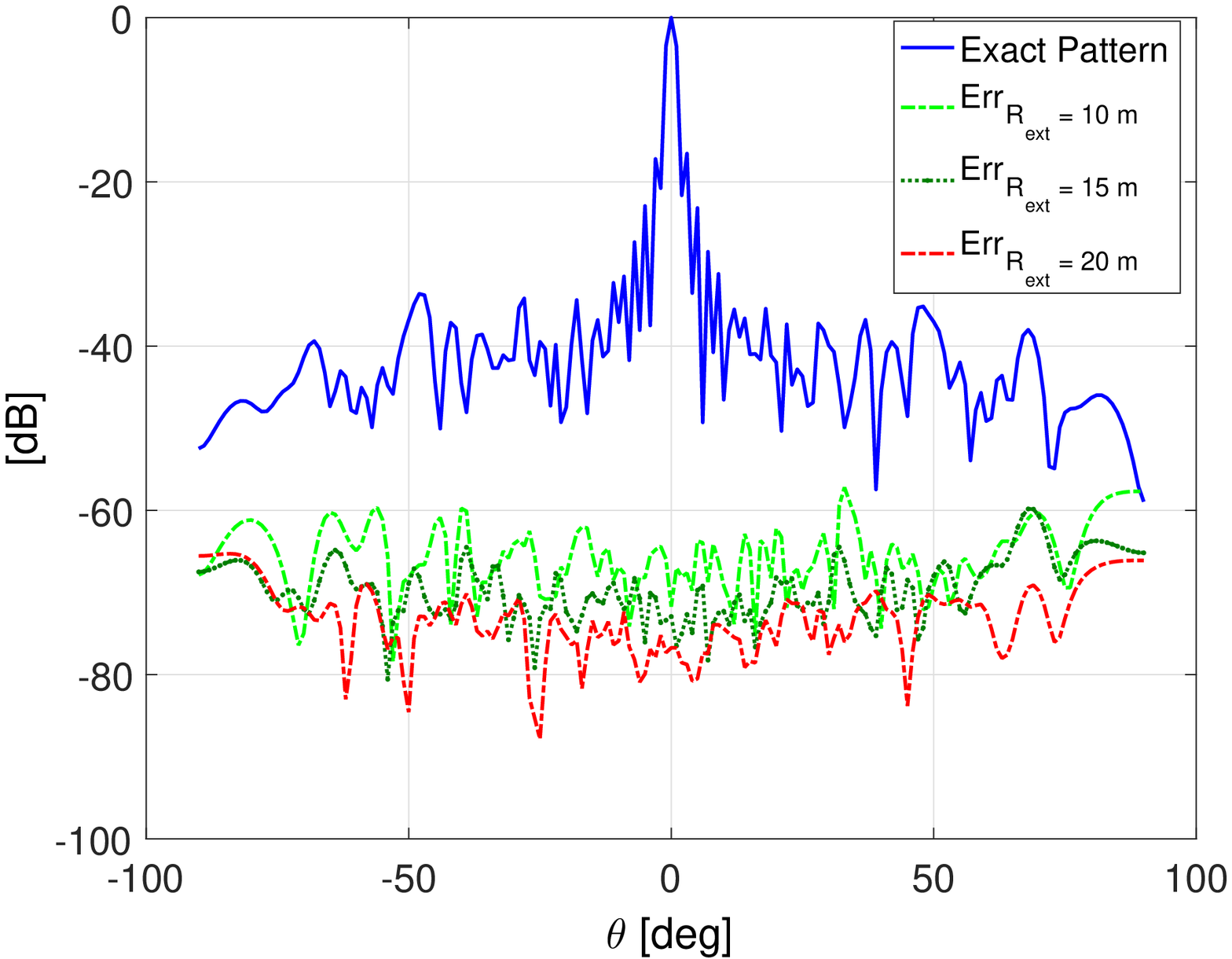}}

\caption{Normalized radiation pattern of the array in Fig.~\ref{fig:array1500} at 110\,MHz and errors using the tessellation approach in two principal planes. (a) E-plane, and (b) H-plane.}
\label{fig:subArray1500}
\end{figure*}

\section{Simulation of Extremely Large Arrays}  \label{sec:largeArray}

This section shows the performance of the sparse matrix and tessellation techniques on the analysis of very large arrays. The impact of the Radius of Influence (RI) on the array patterns is studied on arrays of different sizes containing 256, 500, 800, 1000 and 1500 elements. The latter has a diameter of 85\,meters, as shown in Fig.~\ref{fig:array1500}, and the same array density has been considered for all the other arrays. The arrays are simulated at 110\,MHz using the HARP approach; EEPs and the array pattern are calculated. 
It is noted that the  array of 1500 elements cannot be handled using CST nor WIPL-D with our current hardware.


\begin{table}
\centering
  \begin{threeparttable}
    \renewcommand{\arraystretch}{1.3}
		\caption{computational Complexity by Sparse Matrix and Tessellation approaches.}
		\label{tab:comptLargeArray}
     \begin{tabular}{c c c}
			\hline
			& \textbf{Sparse Matrix} & \textbf{ Tessellation}\\
			\hline
			\hline
			Filling Matrix & $\mathcal{O}\Big(N_a N_I N_m^2$\Big) &  $\mathcal{O}\Big(N_a N_s N_m^2\Big)$ \\
			\hline
			Solving & $\mathcal{O}\Big(N_a N_I^2 N_m^3\Big) $ & $\mathcal{O}\Big(N_a N_s^2 N_m^3\Big)$ \\
			\hline
			\end{tabular}
    \begin{tablenotes}
      \footnotesize
      \item where $N_I$ and $N_s$ are average number of elements within the RI and the subarray, respectively.
    \end{tablenotes}
  \end{threeparttable}
\end{table}

\subsection{Sparse Matrix Approach}

Fig.~\ref{fig:array1500RI} shows the exact patterns and the error for different RIs in two main planes for the 1500 elements array. The errors are evaluated using (\ref{eq:eep_error}), where $\overrightarrow{E}(\theta,\phi)$ and $\overrightarrow{E}^{\textrm{ref}}(\theta,\phi)$ stand for the exact pattern and those obtained using the sparse array approach, respectively. It is clear that for larger values of the RI, the errors are smaller, since more elements are included in the RI region. It is also seen that, using only the isolated pattern, the error level is higher than the sidelobe level at far-out angles. Even for RI = 30\,m, the error is close to some sidelobes. The results again show the strong effect of mutual coupling in a random array made of SKALA elements. Fig.~\ref{fig:RIconvergence} displays the convergence of error curves versus the RI for different array sizes at 110\,MHz. It shows the trend of reducing errors for larger RI and one can observe that to achieve a given error level, larger arrays require a smaller RI (normalized to array size). This is probably due to the slow averaging effect of randomization of the array on mutual coupling. As the RI is smaller (w.r.t. array size) for large arrays, the interaction matrix will be sparser, which implies the capability to analyze large arrays using the RI concept. The computational complexity of the approach is detailed in Table~\ref{tab:comptLargeArray}, where $N_I$ is the average number of elements within the RI, which is proportional to $\rm{RI}^2$.

\subsection{Tessellation Approach}

Fig.~\ref{fig:subArray1500} shows the radiation patterns and errors using the tessellation approach on the array of 1500 SKALA elements shown in Fig.~\ref{fig:array1500} for $\rm{R_{in}} = 10\,m$ and different $\rm{R_{ext}}$ values. It is interesting to see that tessellation is able to provide high-quality solutions w.r.t. the exact ones as the error levels are quite low even for small $\rm{R_{ext}}$. Despite sharing the same RI idea as the sparse array approach, the tessellation seems  to perform better. The reason lies in the fact that even if the sparse array approach only accounts for the interaction between elements inside the RI, the solution of the resulting system of equations still provides non-zero currents for elements outside of the RI region, while the excited elements remain within the RI. This ``artificial'' current contributes to the radiation patterns, while probably not being highly relevant nor accurate. Conversely, the tessellation approach only solves for the subarray and calculates the contribution of those elements inside the subarray.

\begin{figure}[!htb]\centering
\includegraphics[scale=0.475,clip,trim={0cm 0cm 0cm 0cm}]{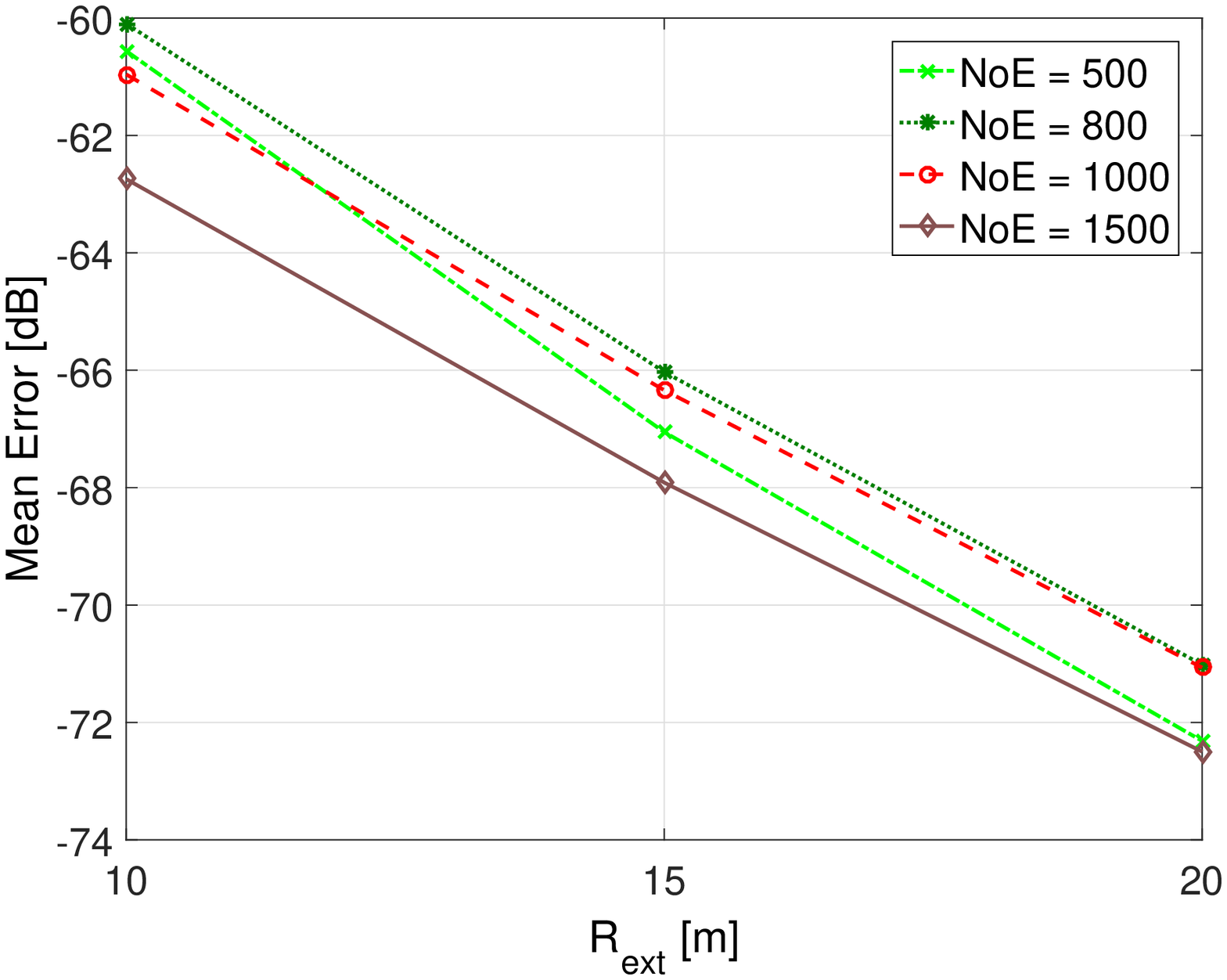}
\caption{Mean errors vs. $\rm{R_{ext}}$ for different array sizes using the tessellation approach at 110\,MHz.}
\label{fig:subArrayConvergence}
\end{figure}

Fig.~\ref{fig:subArrayConvergence} shows the convergence of the tessellation approach for different array sizes. As $\rm{R_{ext}}$ increases, the error decreases almost linearly in logarithmic scale.  This allows one to obtain a given accuracy by finding a proper $\rm{R_{ext}}$. Increasing $\rm{R_{ext}}$, however, results in increasing the computational complexity, both for matrix filling and for solving the system of equations. An optimum choice of $\rm{R_{ext}}$ can be found following Eq.~(\ref{eq:complexity}), as discussed in Section~\ref{subsec:LargeArray}; it favors an extended region comprising half of all neighboring inner-hexagons.  A detailed complexity of the tessellation approach is listed in Table~\ref{tab:comptLargeArray}, where $N_s$ is the average number of elements in a subarray, which is proportional to $\rm{(R_{in} + R_{ext})^2}$ with $\rm{R_{in} = R_{ext}}$.
By selecting $N_I$ and $N_s$, i.e. choosing $\rm{RI}$ and $\rm{R_{in}}$, respectively, one can obtain a same complexity for both methods. For extremely large arrays, $N_I$ and $N_s$ are at least one order of magnitude smaller than $N_a$, which results in a big saving factor compared to the direct solution over the whole array.

It is also interesting to see that the convergence curve in Fig.~\ref{fig:subArrayConvergence} is quite independent from the array size, which assures the performance of the tessellation approach for other (larger) arrays, and allows one to select $\rm{R_{in}}$ and $\rm{R_{ext}}$ in advance.
Regarding the analysis, the tessellation approach offers the possibility of parallelizing the computation as the analysis of each subarray is totally independent. Moreover, the tessellation is not limited by the array size as each subarray can be solved effectively with a present-day desktop computer, providing enough temporary memory RAM to store the reduced MoM matrix and solutions, for example 16\,GB has been proven sufficient in all tested cases. For the largest array, i.e. consisting of 1500 SKALA antennas, the analysis took about 15 minutes using the tessellation approach to simulate the array and calculate all the EEPs.


\section{Conclusion} \label{sec:conclusion}

A fast full-wave technique for the analysis of large arrays made of complex 3-D metallic antennas has been presented. The calculation of interactions between antennas has been accelerated using the HARP technique, where a non-expensive model is built for MBF interactions. The reduced matrix is then rapidly filled using such model. Several improvements have been added to HARP to cope with the complexity of solving very large arrays. An experimental example of compact array and simulations of SKALA arrays have demonstrated the performance of HARP. Furthermore, a tessellation approach has been proposed for analyzing very large arrays of SKALA elements, offering a solution to the analysis of extremely large arrays,  such as the SKA telescope, for which hundreds of stations are foreseen. In particular, the proposed approach enables the analysis of very large arrays on present-day desktop computers, while providing highly accurate results. 

The simulation technique opens the possibility to integrate the full-wave simulation in an optimization procedure~\cite{ClavierTAP14}, which is capable of taking into account mutual coupling to optimize the performance of the arrays (e.g. in terms of sensitivity, including noise coupling). Moreover, the accurate EEPs obtained using HARP can be incorporated in the calibration stage for modern radars and astronomical instruments such as the SKA. The use of the PMCHWT approach~\cite{Harrington01}, combined with the surface equivalence principle~\cite{CCRadio09}, allows the extension of HARP to antennas comprising dielectric material.

\section*{Acknowledgment}
The authors would like to thank the UK's STFC funding that has supported this work. They are also grateful to their SKA colleagues for numerous discussions about SKA station modeling.

\ifCLASSOPTIONcaptionsoff
  \newpage
\fi

\end{document}